\documentclass{article}

\usepackage{emulateapj}
\usepackage{apjfonts}
\usepackage{graphics}

\newenvironment{inlinefigure}{%
\def\@captype{figure}%
\noindent\begin{minipage}{0.999\linewidth}\begin{center}}
{\end{center}\end{minipage}\smallskip}

\newlength{\colwidth}
\newcommand{\HI}{\ion{H}{1}}

\newcommand{\OVI}{\ion{O}{6}} 
 
\newcommand{\SiIV}{\ion{Si}{4}}
\newcommand{\MgII}{\ion{Mg}{2}}
\newcommand{\simgt}{\ga}
\newcommand{\simlt}{\la}
\newcommand{\K}{{\rm K}}

\lefthead{Carswell et al.}
\righthead{\ion{O}{6} at $z\sim 2$}

\begin{document}


\title{The Enrichment History of the Intergalactic Medium:\\
\ion{O}{6} in Ly$\alpha$ forest systems at redshift $z\sim 2$\altaffilmark{1}}

\author{Bob Carswell\altaffilmark{2}, Joop Schaye\altaffilmark{3},
and Tae-Sun
Kim\altaffilmark{4}}

\altaffiltext{1}{The data used in this study are based on public data from UVES Commissioning and Science verification at
the VLT/Kueyen telescope, ESO, Paranal, Chile.}

\altaffiltext{2} {Institute of Astronomy, Madingley Road, Cambridge CB3
0HA, UK}

\altaffiltext{3} {School of Natural Sciences, Institute for Advanced
Study, Einstein Drive, Princeton NJ 08540}

\altaffiltext{4} {European Southern Observatory, Karl-Schwarzschild-Stra{\ss}e 2, D-85748 Garching bei
M\"unchen, Germany}

\begin{abstract}
A search for \ion{O}{6}\,1031.93,\,1037.62\,\AA\ at redshifts
corresponding to Ly$\alpha$ lines in the $z_{\mathrm{em}}\sim 2.4$ QSOs ${\rm
HE}1122-1648$ and ${\rm HE}2217-2818$ reveals that a substantial
fraction of those with \ion{H}{1} column densities $\log
N({\rm HI}) >14.0$ (cm$^{-2}$) are highly ionized and show some heavy
element enrichment.  If these two sight lines are typical, then the \ion{O}{6}
systems contain a cosmologically significant fraction of the baryons
and the metals in the universe. 
For most systems 
the temperatures derived from the line
widths are too low for collisional ionization to be responsible for the
\ion{O}{6} lines. Photoionization models with a substantial hard ultraviolet
flux can reproduce the observations for densities that are in good
agreement with a model assuming local, hydrostatic equilibrium and
heavy element abundances in the range $\sim 10^{-3}$-$10^{-2}$
solar. 
Photoionization by a UV flux much softer than that predicted by
Haardt \& Madau (1996) for a background dominated by quasars can be
ruled out. Finally, we find one system with a very low \HI\ column
density for which both photoionization
and collisional ionization models yield a metallicity close to solar and a
density that is inconsistent with gravitational confinement, unless the 
gas fraction is negligible.  
\end{abstract}


\section{Introduction}

There have been a number of searches for heavy elements in the low
density intergalactic medium over a range of redshifts, motivated in
part by the need to investigate possible enrichment mechanisms and to
investigate the spectrum of the integrated ionizing background. The
intergalactic medium is observable through the presence of Ly$\alpha$
and heavy metal absorption lines in QSO spectra at redshifts up to that
of the background QSO.  Cowie et al.\ (1995) showed that for many of
the high \ion{H}{1} column density systems ($N_{\rm HI}\simgt
10^{14.5}$~cm$^{-2}$) \ion{C}{4} is also present.  Cowie \& Songaila (1998) and
Ellison et al.\ (2000) extended this analysis by a pixel optical depth
technique to find that a similar \ion{C}{4}/\ion{H}{1} ratio applies at lower column
densities at redshifts $z\sim 3$. The redshift dependence of the
\SiIV/\ion{C}{4} column density ratio \cite{Songaila96} suggests that there is
a change in the nature of the ionizing background at $z\sim 3$, though
a further investigation provides little evidence for any redshift
dependence \cite{Boksy97}.

While these early investigations concentrated on \ion{C}{4}~1548.20,~1550.78
since it has the strongest lines falling in the spectral region clear
of contamination from the stronger Lyman forest absorption, at higher
redshift 
photoionized
\ion{O}{6}~1031.93,~1037.62 is expected to be a much better probe of
the heavy element enrichment
in the low density intergalactic medium
(Chaffee et al., 1986;
Rauch, Haehnelt \& Steinmetz, 1997).  However, at high redshifts the
presence of Lyman forest lines makes the detection of the \ion{O}{6} lines
very difficult.  A thorough search by Dav\'e et al.\ (1998) at
redshifts $\simgt 3$ did not reveal any convincing candidates. However,
Schaye et al.\ (2000) have shown using a pixel optical depth technique
that \ion{O}{6} is generally present in the Lyman forest absorption systems
seen in quasar spectra at redshifts $\simlt 3$.  \ion{O}{6} absorption systems
at low redshifts ($z \simlt 1$) are common (Burles \& Tytler 1996;
Tripp, Savage \& Jenkins, 2000), and a small number of such
identifications have been found towards the $z=1.73$ QSO HE0515-4414
\cite{Reimers01}.

At redshifts $z\simgt 2$ the \ion{O}{6} lines are in a wavelength
region where they are measurable using ground-based facilities, and at
$z\sim 2$ the Lyman forest confusion is less than at higher redshifts.
Therefore, it should be easier to detect the \ion{O}{6} lines at $z\sim
2$ than at higher redshifts.  Here we describe the results of a search
for identifiable lines from the \ion{O}{6}$\,1031.93,\,1037.62$ doublet
in specific absorption systems towards the $z=2.412$ QSO ${\rm
HE}1122-1648$ and the $z=2.406$ QSO ${\rm HE}2217-2818$.  We
investigate heavy element abundances and the nature of the ionizing
mechanism for some representative cases where the \ion{H}{1},
\ion{C}{4} and \ion{O}{6} lines could arise in the same regions. There
are also systems similar to those reported by Simcoe et al. (2002),
where there are clear velocity differences between these ions.

\section{Observations and absorption line fits}

The data reduction and analysis for HE$2217-2818$ have been described
by Kim, Cristiani \& D'Odorico (2001), who also show part of the
spectrum.  Some spectral regions and results for HE$1122-1648$ are
described by Kim et al.\ (2002). The spectral range covered by the
analysis is from 3050\,\AA\ to the \ion{C}{4} emission line in both cases.

The continuum was initially determined as described by Kim et al.\
(2002), and the absorption lines fitted by multiple Voigt profiles,
using the VPFIT program described by Webb (1987), Rauch et al.\ (1992) as
updated by Carswell et al.\ \\
(see
http://www.ast.cam.ac.uk/$_{^{\sim}}$rfc/vpfit.html). 
The whole of the available spectrum was fitted so as to provide a
self-consistent set of line profile parameters for each ion. Except
where noted otherwise below, different ions with components in the same
system were not constrained to have the same redshift, and their
Doppler parameters were fitted independently. The main exceptions are
the systems containing \ion{O}{6} described below where ionization
levels were investigated. Each ion was fitted using all observed
transitions which provided useful information.

Heavy element lines at wavelengths above the Ly$\alpha$ emission line
were fitted first, including components in the Ly$\alpha$ forest region
where necessary, and were used as indicators of where other ions might
appear in the forest. Since we are interested in
\ion{O}{6}\,1031.93,~1037.62 over a range of redshifts, we have to be
aware not only of the possibility of contamination by Ly$\alpha$
absorption at lower redshifts, but also by Ly$\beta$, $\gamma$ etc.
Furthermore, these contaminating lines affect not only the \ion{O}{6}
lines themselves, but also the continuum around them. The strategy we
adopted to allow for this, was to fit the Lyman lines 
in decreasing redshift order, taking care to include all previously
fitted higher order Lyman lines when moving down in redshift.  This
procedure is generally satisfactory, but in a few cases a putative low
redshift Ly$\alpha$ introduced to obtain a satisfactory fit to a higher
order Lyman line at some redshift turns out to be predominantly a
different higher order Lyman line from another redshift system. In such
cases the data were refitted with the misidentified Ly$\alpha$ removed or
modified.  The net result after this iterative process is a set of
Voigt profiles which fits the entire Lyman forest
spectrum.  Attempting to measure the \ion{O}{6} lines without doing
this, and effectively using only \ion{O}{6} and Ly$\alpha$, would give
misleading results in many cases.

Fitting different transitions, particularly ones which are widely
spaced in wavelength, revealed that in many cases the adopted continuum
levels were not precisely correct, and local adjustments were needed to
obtain a self-consistent fit. These continuum adjustments were made as
part of the minimum $\chi^2$ parameter optimization within the VPFIT
program.  With few exceptions these continuum changes were less than
1\% of the originally adopted value, and tended to be correlated
between neighboring fitting regions used. In light of this, a new
continuum was produced by interpolating over the continuum deviations
to provide a smooth, slowly-varying curve, which was then used to
adjust the original continuum.  The whole spectrum was then refitted
without allowing the continuum to vary to produce the final set of
Voigt profile parameters.

A sample list of the line fit parameters obtained in this way for
HE$1122-1648$ is given in Table \ref{tab:sample}. The table gives the
ion, redshift $z$, Doppler parameter $b$ in km~s$^{-1}$ and log column
density $N$ (cm$^{-2}$) for each system found. The error estimates
given against each quantity are from the diagonal terms of the
covariance matrix from the fitting procedure, and correspond to
$1\sigma$ estimates if the variables are uncorrelated. Where no error
estimate is given, the variable involved has been tied to others on the
same system.  For the redshift $z=2.101764$ example in the table, the
redshifts of \ion{H}{1}, \ion{C}{4} and \ion{O}{6} have been
constrained to be the same, and the Doppler parameters have been
constrained so that $b^2 = b_{\rm turb}^2 + b_{\rm therm}^2$, where
$b_{\rm turb}$ is identical for all ions and $b_{\rm therm}^2 = 2kT/m$,
where $m$ is the ion mass and the temperature $T$ is constrained to be
identical for all ions. The quantity $b_{\rm turb}$ is not necessarily
true turbulence, but bulk motion, such as the Hubble expansion of the
absorbing regions, may be approximated in this way. We choose not to
give observed wavelengths for the lines, since more than one transition
from an ion was usually included in determining these parameters. The
full line lists for this object and HE2217--2818 are available as part
of the electronic journal article.

\section{O\,VI systems}

The fitted parameters for the systems where high ionization species may
be present are given in Tables \ref{tab:OVI2217} and \ref{tab:OVI1122}.
For HE1122--1648 the redshift range searched is $2.000<z<2.365$ and for
HE2217--2818 $2.000<z<2.360$. At redshifts $z<2.0$ the signal-to-noise
ratio in the regions of the spectra containing the \ion{O}{6} doublet
is too low to allow meaningful measurements.  We have imposed a high
redshift cutoff at a redshift 4000 km~s$^{-1}$ below the emission
redshift to avoid the region near the background QSO.

There are a number of ways in which absorption lines from different
ions, which are thought to arise in the same system, can be fitted with
Voigt profiles: the redshifts can either be constrained to be the same
or they can be allowed to vary independently, and the Doppler
parameters can either be determined independently or be constrained to
be consistent with a picture where the line is broadened by a mixture
of turbulent and thermal motions. Since we are interested in ionization
levels in single regions, we have chosen to constrain the redshifts to
be identical and the Doppler parameters to consist of a turbulent and a
thermal component as described in the previous section. Note that we
implicitly assumed that all bulk/turbulent motions have a Gaussian
velocity distribution.

The $1\sigma$ error estimates for the temperatures listed in Tables
\ref{tab:OVI2217} and \ref{tab:OVI1122} are determined using an
approximation based on small standard deviation normal distributions
for the Doppler parameters, so in some cases negative temperatures
appear to be in the allowed range when turbulent velocities
dominate. In the tables the redshift is left blank if it is
constrained to be the same as that of the ion above it in the
list. The Doppler parameters are given for all detected ions of each
redshift component, with individual errors only for one ion as a
guide, even though all line widths were included in determining these.

The \ion{O}{6}\,1031.93,~1037.62 lines are invariably in the Ly$\alpha$
forest and are usually blended with higher order Lyman lines from
higher redshift systems, so any putative identifications rely strongly
on the presence of both lines with the correct doublet ratio. In many
cases, as described individually below, the blending is dominated not
by Ly$\alpha$, but by higher order Lyman lines or by some other
species.  In such cases the parameters of the contaminating lines are
constrained by fits to other absorption systems, and the measured
strength of the \ion{O}{6} component is therefore more robust than if
the contamination were dominated by Ly$\alpha$.  Of course it is still
likely that in some cases the \ion{O}{6} identification may be
incorrect, with a chance superposition of Lyman lines and/or noise
giving the appearance of the \ion{O}{6} doublet. Against this, there is
a bias toward finding components with Doppler parameters less than
$\sim 15$ km~s$^{-1}$, since broader ones are more easily confused with
Lyman forest systems. Our approach has been to try \ion{O}{6} in all
possible cases, and retain it as a reasonably firm identification when
a physically self-consistent fit can be obtained.  The search for
\ion{O}{6} was performed without regard to the strength of the
corresponding Ly$\alpha$ line, since the neutral hydrogen fraction
could be small.

We can obtain an estimate of the number of spurious \ion{O}{6}
identifications by using wavelengths offset from the real
ones. A fake doublet with rest wavelengths 1035.0 and 1041.0 was chosen
arbitrarily, but close to the real \ion{O}{6} doublet wavelengths so
that the signal-to-noise ratio in the data is similar.  A search for
fake \ion{O}{6} corresponding to \ion{H}{1} with column densities
$>10^{14}$~cm$^{-2}$ in HE$1122-1648$ revealed that in 8 of the 24
cases the fake ion could be present. In the other 16 cases at least one
of the fake \ion{O}{6} lines was demonstrably absent. This suggests
that up to about 1/3 of the \ion{O}{6} identifications could be
spurious. However, we note that many of our \ion{O}{6} identifications
were then found to have detectable \ion{C}{4} at the same redshift, which
increases the likelihood that they are real. 

It is interesting that for all but 2 of the 8 systems with $\log N({\rm
HI})>14.5$ in HE1122--1648 the data are consistent with the presence of
\ion{O}{6}, and it may even be weakly present in one of the
exceptions.  In contrast, the presence of the fake doublet is excluded
for 5 of the 8 systems.

\subsection{Comments on individual systems}

Below we present an atlas of all the absorption systems in which
\ion{O}{6} could be present, and discuss each system individually.
Even though most of the \ion{O}{6} lines occur in blends, the fits are
much better constrained than the figures suggest because a large
fraction of the absorption arises in higher order Lyman lines for which
the parameters are determined from other hydrogen transitions. The
uncertain cases are those for which there are blends with Ly$\alpha$
components close to the \ion{O}{6} transitions.  The Ly$\alpha$ and
other better-constrained components blended with the \ion{O}{6} lines
are shown in Figures \ref{fig:z2p074} to \ref{fig:z2p352} along with the
profiles for the lines in the individual systems.

\subsubsection{HE2217--2818}

\begin{description}
\item[$z=2.075$] (Fig. \ref{fig:z2p074}):\\ 
Ly$\beta$ shows the two-component nature of the \ion{H}{1} clearly. The
\ion{O}{6} lines at $z=2.075446$ are not strongly blended with Lyman
forest lines, but at $z=2.074740$ both the \ion{O}{6}~1031.93 and
1037.62 lines are seen as sharp components blended with broader
Ly$\beta$ from systems at higher redshift and nearby strong Ly$\alpha$
lines.  \ion{C}{4}\,1548.20,~1550.78 are present in both components,
though in the higher redshift case there is a significant difference
between the \ion{C}{4} and \ion{O}{6} redshifts.

\item[$z=2.180$] (Fig. \ref{fig:z2p180}): \\
This system is probably more complex than the
component structure given in the table would suggest. It shows ions
from \ion{C}{2} to \ion{O}{6} apparently coexisting in one
component.  We have not tried for a physically self-consistent model or
estimated temperatures because of the complex structure. For all
components in the complex the \ion{O}{6} lines are free of any significant
blending from other systems.
\end{description}

\subsubsection{HE1122-1648}

\begin{description}
\item[$z=2.007$] (Fig. \ref{fig:z2p007}): \\
The \ion{C}{4}\,1548.20,~1550.78 lines are clearly
broader than \\
\SiIV\,1393.76,~1402.77, and are best fitted as two components,
one at the \SiIV\ redshift, $z=2.007109$, and another agreeing with \ion{O}{6},
$z=2.007205$.  Both lines in the \ion{O}{6} doublet are in a wavelength region
where the signal-to-noise ratio is low. \ion{O}{6}\,1031.93 is in the wing of
a weak Ly$\theta$ and is otherwise free of significant contamination.
\ion{O}{6}~1037.62 is seen as a distinct sharp component between a Ly$\alpha$
line at $z=1.566566$ and a blend of higher order lines along with Ly$\alpha$
at $z=1.567140$.
\end{description}

\vspace{1cm}
\begin{inlinefigure}
\centerline{\resizebox{0.68\colwidth}{!}{\includegraphics{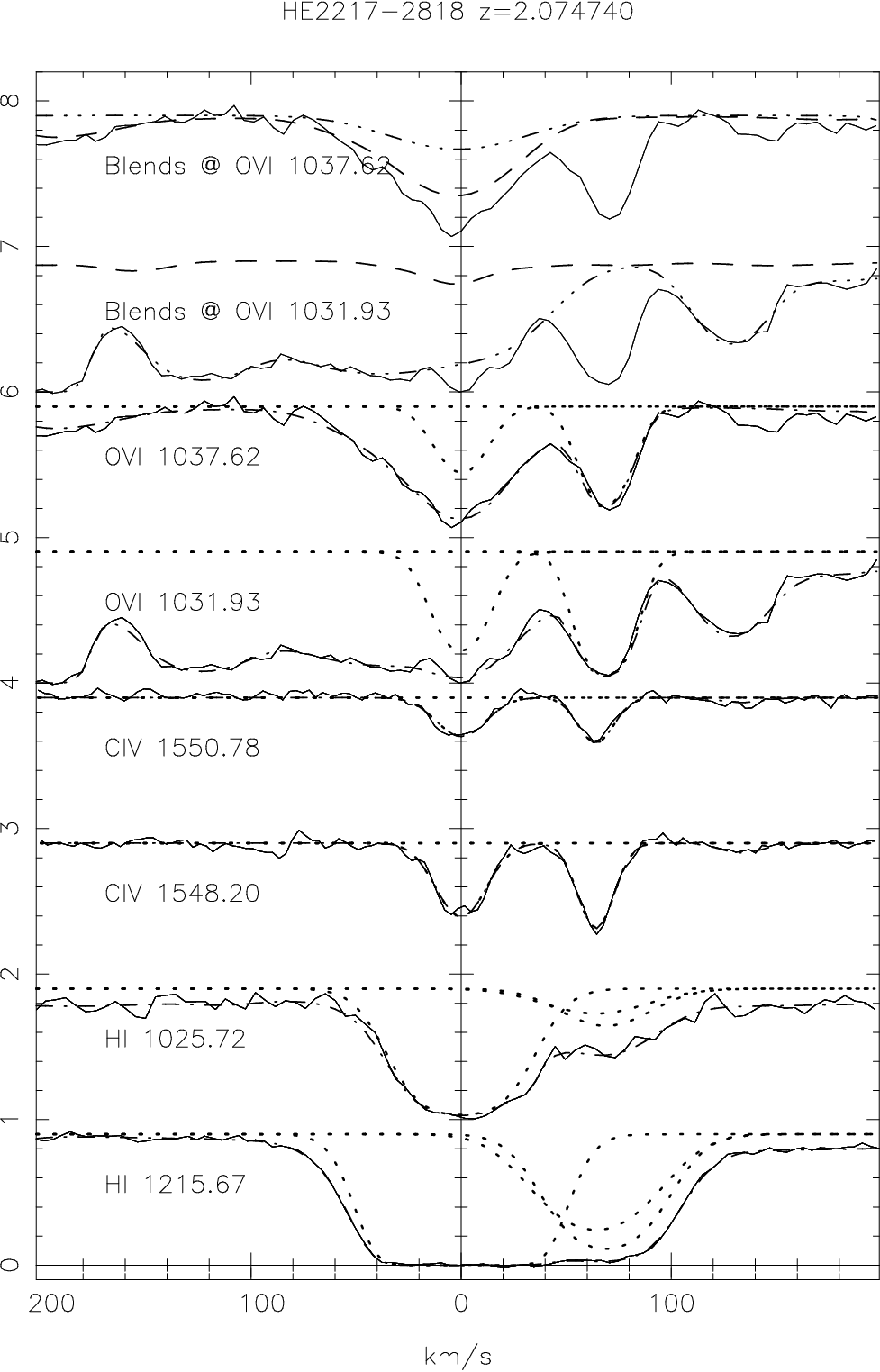}}}
\figcaption[z2p074p.ps]{The systems at $z=2.074740$ and 2.075446 in
HE2217--2818 showing the profile fits to the \ion{H}{1}, \ion{C}{4} and
\ion{O}{6} regions (dot-dashed lines) against the data on a velocity
scale centered on the redshift given above the figure.  For \ion{H}{1}
and \ion{O}{6} each of the spectral regions is normalized to a
continuum value of 0.9, and an integer offset has been applied to
separate them.  In this case \ion{C}{4} scale has been multiplied by a
factor three, so the zero levels for the \ion{C}{4} lines are 2.7 units
below their continuum values. The individual components, here at
$z=2.074740$, 2.075398 and 2.075446, are shown as dotted lines in the
lower six spectral regions.  The upper two spectra show the \ion{O}{6}
line regions with the features constrained by the presence of other
transitions shown as dashed lines, and the blended Ly$\alpha$ as
dot-dot-dot-dash.  This example provides one reasonably clear and one
blended suggested \ion{O}{6} identification.\label{fig:z2p074}}
\end{inlinefigure}

\begin{inlinefigure}
\centerline{\resizebox{0.68\colwidth}{!}{\includegraphics{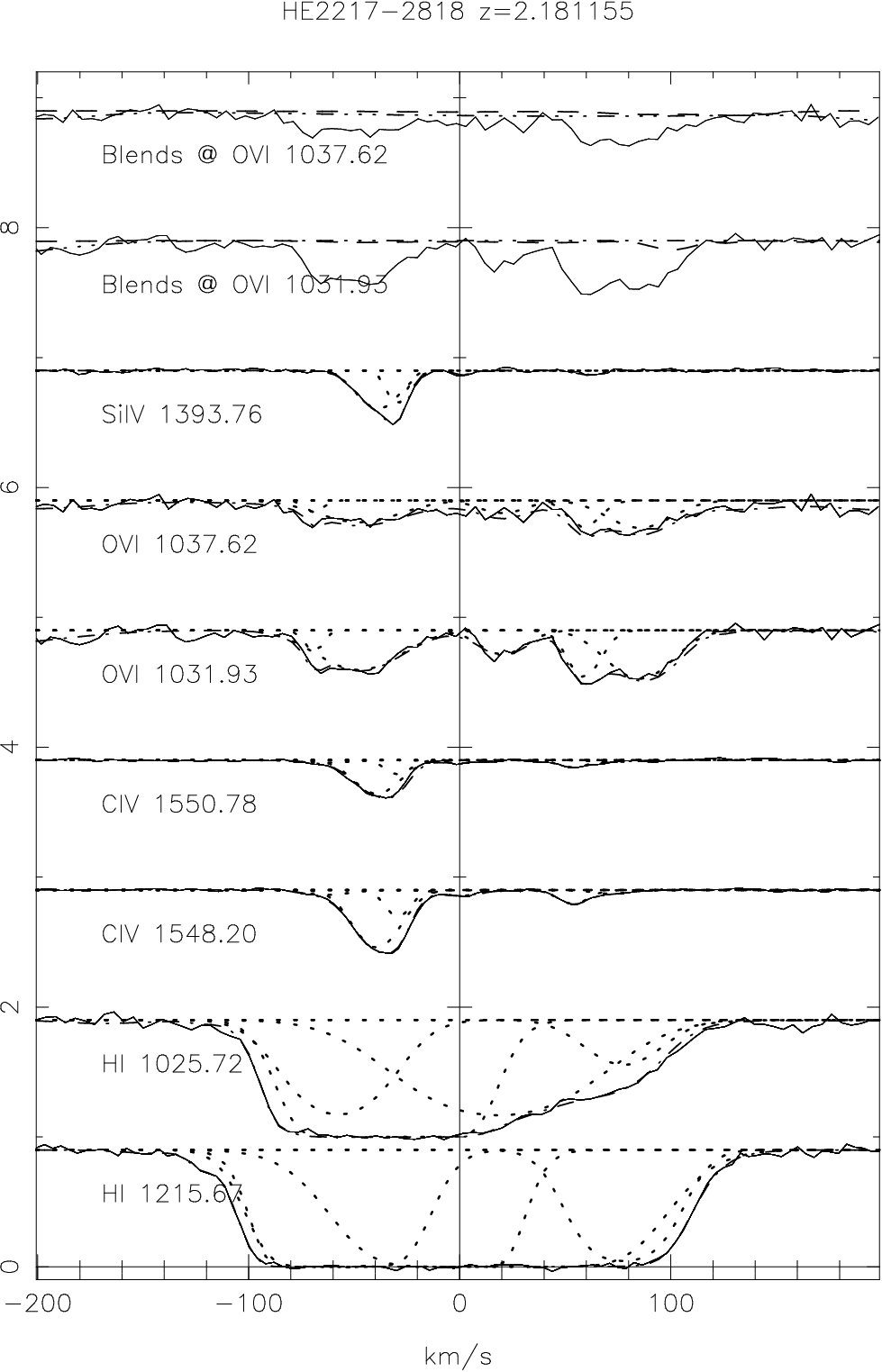}}}
\figcaption[z2p180p.ps]{The systems at $z=2.180$ in HE2217--2818
showing the fitted \ion{H}{1}, \ion{C}{4}, \ion{O}{6} and \ion{Si}{4}
profiles. The \ion{C}{4} region has not been rescaled, and otherwise
the details are as for Fig. \ref{fig:z2p074}. \label{fig:z2p180}}
\end{inlinefigure}

\begin{description}
\item[$z=2.030$] (Fig. \ref{fig:z2p030}):  \\
This system is unusual in
that it shows \ion{C}{4} and NV~1238.82,~1242.80, but only weak
Ly$\alpha$.  Both lines of the \ion{O}{6} doublet are noisy and
blended, and a marginally satisfactory fit can be obtained without
their presence. \ion{O}{6}~1031.93 is in the wing of a blend of
Ly$\beta$ at $z=2.048614$ and Ly$\gamma$ at $z=2.215271$, and
\ion{O}{6}~1037.62 is in a blend of several higher order Lyman lines.
There are no Ly$\alpha$ lines near either component.

\item[$z=2.033$] (Fig. \ref{fig:z2p033}): \\
There is a noise spike
which obliterates the \ion{O}{6} 1037.62 line, so the \ion{O}{6}
identification at $z=2.033133$, and parameter estimates, are based
solely on a clean \ion{O}{6}~1031.93 line and its redshift coincidence
with \ion{H}{1}.  \ion{C}{4}~1548.20 is not seen.

\item[$z=2.064$] (Fig. \ref{fig:z2p064}):\\
The Ly$\alpha$ profile is complex, and three components are present in
\ion{C}{4}~1548.20,~1550.78. \ion{O}{6} is present in the $z=2.065681$ component. The
1031.93 line is blended with Ly$\beta$ at $z=2.084275$. The \ion{O}{6}~1037.62
line is in a blend containing Ly$\alpha$ at $z=1.616531$ and several
higher order Lyman lines from higher redshift systems.

\item[$z=2.080$] (Fig. \ref{fig:z2p080}): \\
If the \ion{H}{1} and \ion{O}{6} line widths
are internally consistent, and the apparent double structure in each of
the \ion{O}{6} lines is real, there appear to be at least two components, at
$z=2.080162$ and 2.080527. In a two-component fit each has a low
estimated temperature but large turbulent line widths. The dominant
blend with the \ion{O}{6}~1031.93 line is Ly$\delta$ at $z=2.346958$, with
parameters well determined from other lines in the Lyman series. The
\ion{O}{6}~1037.62 components fall between Ly$\beta$ at $z=2.115206$ and at
$z=2.116542$, and the wings of these lines affect the apparent
continuum level against which the \ion{O}{6} line is seen. The blue wing of
Ly$\alpha$ at $z=1.629738$ with $b=25.8$ and $\log N({\rm HI})=12.75$
also has a small effect the \ion{O}{6}~1037.62 line profiles.

It is also possible to fit this system as a single component, with the
\ion{O}{6} lines having a Doppler parameter of 39 km~s$^{-1}$ and \ion{H}{1}
with a lower $b$ ($= 29$ km~s$^{-1}$), though it is not clear what
this means. In this case $\log N({\rm HI})=15.1$ and $\log
N({\rm OVI})=13.9$.

\item[$z=2.102$] (Fig. \ref{fig:z2p101}): \\
The structure
around the \ion{O}{6}\,1031.93 feature is dominated by Ly$\delta$ at
$z=2.369161$ shortward and \ion{C}{4} 1548.20 at $z=1.067739$ longward
of the observed wavelength of the line. Ly$\alpha$ is present at
$z=1.632438$, but with $\log N({\rm HI})=13.07$ and $b=49.7$ this does
not strongly affect the \ion{O}{6} line. The \ion{O}{6}\,1037.62 line
is significantly blended with Ly$\alpha$ at $z=1.647796$  ($\log N({\rm
HI})=13.70$, $b=28.1$), and shows only as an extended blue wing to a
feature at 3218.8\AA.

\item[$z=2.206$] (Fig. \ref{fig:z2p206}):\\
\ion{C}{4} is
weak, and the 1548.20 line may have a further velocity component.
\ion{O}{6}~1031.93 is in the wing of a Ly$\alpha$ line at $z=1.721641$,
and 1037.62 with Ly$\alpha$ at $z=1.736637$.

\item[$z=2.215$] (Fig. \ref{fig:z2p215}): \\Two components are detected,
one with \ion{O}{6} line Doppler parameter $b=7.3$~km~s$^{-1}$ at
$z=2.215271$ and a more doubtful one at $z= 2.215494$ with broad lines.
\ion{O}{6}~1031.93 at $z=2.215271$ appears as a narrow component
blended with Ly$\alpha$ at $z=1.729154$ ($b=35.9$, $\log N({\rm
HI})=12.93$), and for 1037.62 the higher redshift component is blended
with Ly$\alpha$ from a system with $z=1.744575$, $b=25.0$ and $\log
N({\rm HI})=12.31$.

\item[$z=2.339$] (Fig. \ref{fig:z2p339}): \\ In this case the presence
of \ion{O}{6} rests solely on a weak 1031.93 line at $z=2.339260$.
\ion{O}{6} 1037.62 at this redshift falls close to a strong \ion{C}{4}
1550.78 at $z=1.234266$. The corresponding wavelength of Ly$\alpha$ is
significantly offset from the main component, and the \ion{H}{1} column density
at $z=2.339260$ is very uncertain. The strongest neighboring \ion{H}{1} components are included in Table \ref{tab:OVI1122}.

\item[$z=2.352$] (Fig. \ref{fig:z2p352}):\\
Two components are
measured at $z=2.352160$ and 2.352695. At each redshift the
\ion{O}{6}~1037.62 line is sufficiently far from the nearest
contaminant, a weak Ly$\alpha$ in each case, that blending does not
significantly affect them. At the lower redshift the \ion{O}{6}~1031.93
line appears as an extended wing of \ion{C}{4}~1548.20 at $z=1.234266$.
From the corresponding \ion{C}{4} 1550.78 it is clear that this
extended wing cannot be due to \ion{C}{4}, even if the blended
\ion{O}{6}~1037.62 at $z=2.339257$ is absent there. At $z=2.352695$
\ion{O}{6}~1031.93 is blended with Ly$\alpha$ at $z=1.846002$
($b=18.1$, $\log N({\rm HI})=13.00$).
\end{description}

\vspace{1cm}
\begin{inlinefigure}
\centerline{\resizebox{0.75\colwidth}{!}{\includegraphics{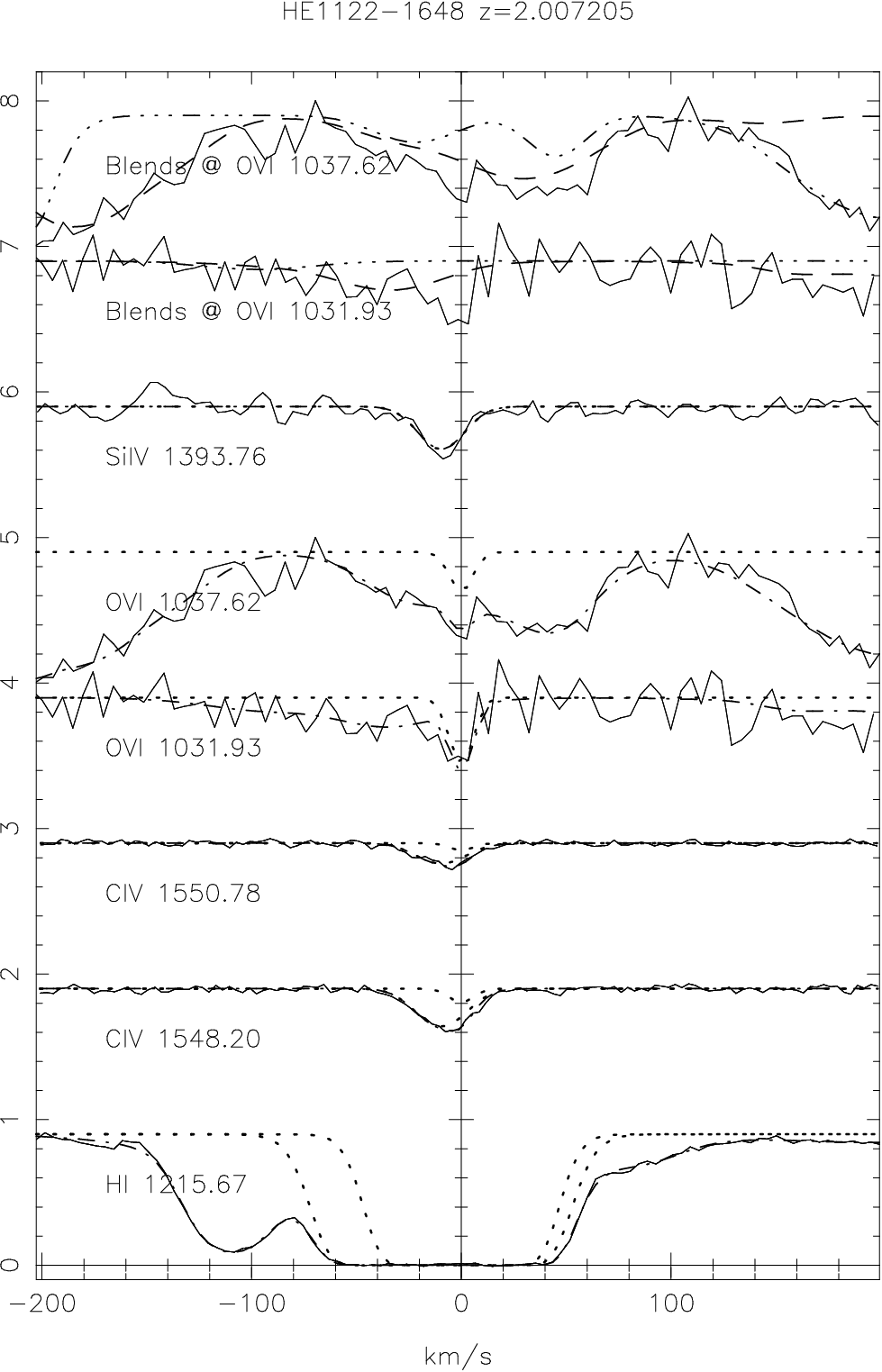}}}
\figcaption[z2p007p.ps]{ The systems at $z=2.007109$ and $2.007205$ in
HE1122--1648 showing the fitted \ion{H}{1}, \ion{C}{4}, \ion{O}{6} and
\ion{Si}{4} profiles. The \ion{Si}{4} region has been scaled by a
factor five. Other details are as for Fig. \ref{fig:z2p074}.
\label{fig:z2p007}}
\end{inlinefigure}

\begin{inlinefigure}
\centerline{\resizebox{0.75\colwidth}{!}{\includegraphics{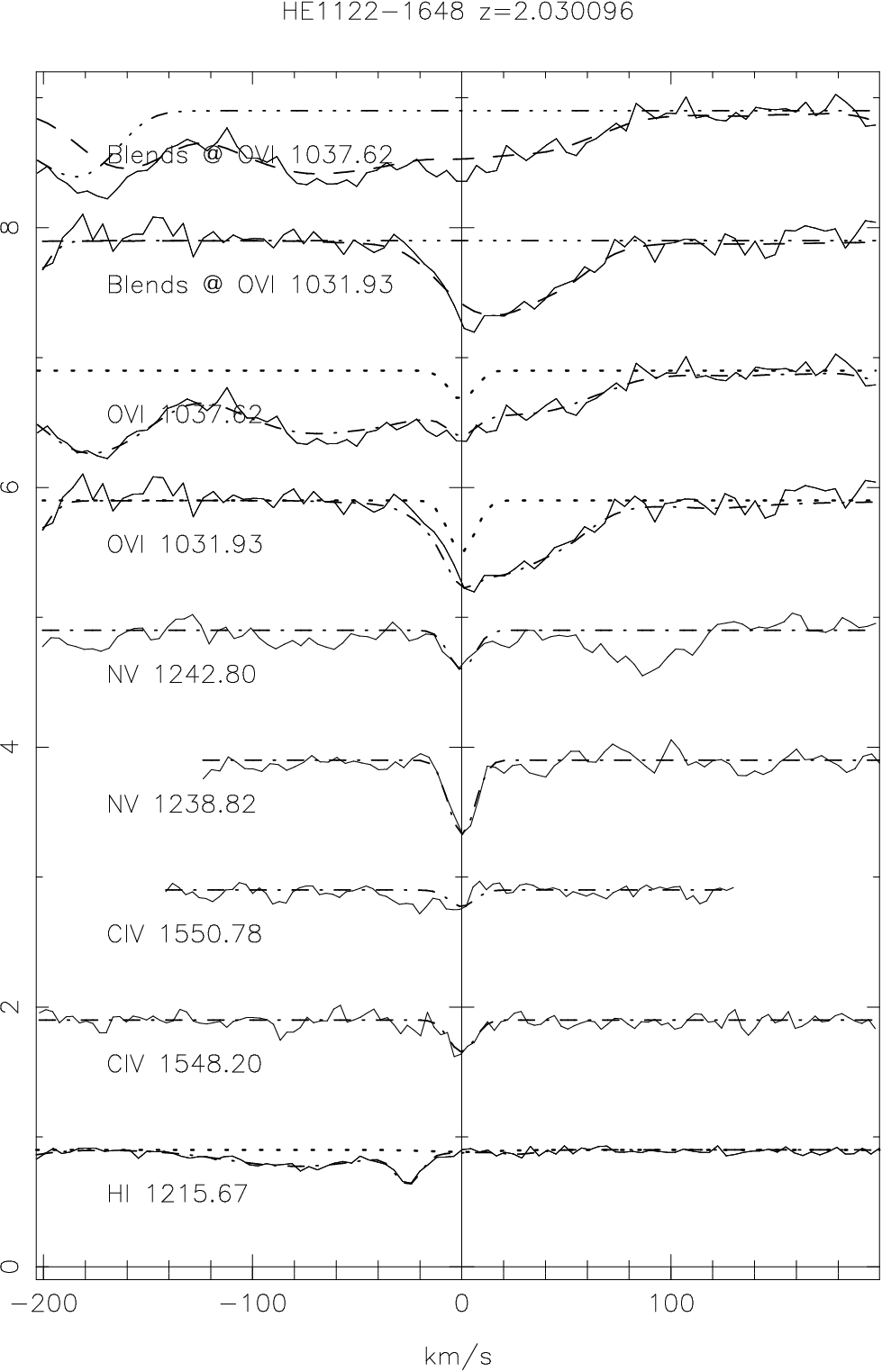}}}
\figcaption[z2p030p.ps]{ The system at $z=2.030096$ in HE1122--1648
showing the fitted profiles in the regions of the \ion{H}{1},
\ion{C}{4}, \ion{N}{5} and \ion{O}{6} absorption lines. The \ion{C}{4}
and \ion{N}{5} regions have been scaled by a factor three. Other
details are as for Fig. \ref{fig:z2p074}.  \label{fig:z2p030}}
\end{inlinefigure}

\begin{inlinefigure}
\centerline{\resizebox{0.75\colwidth}{!}{\includegraphics{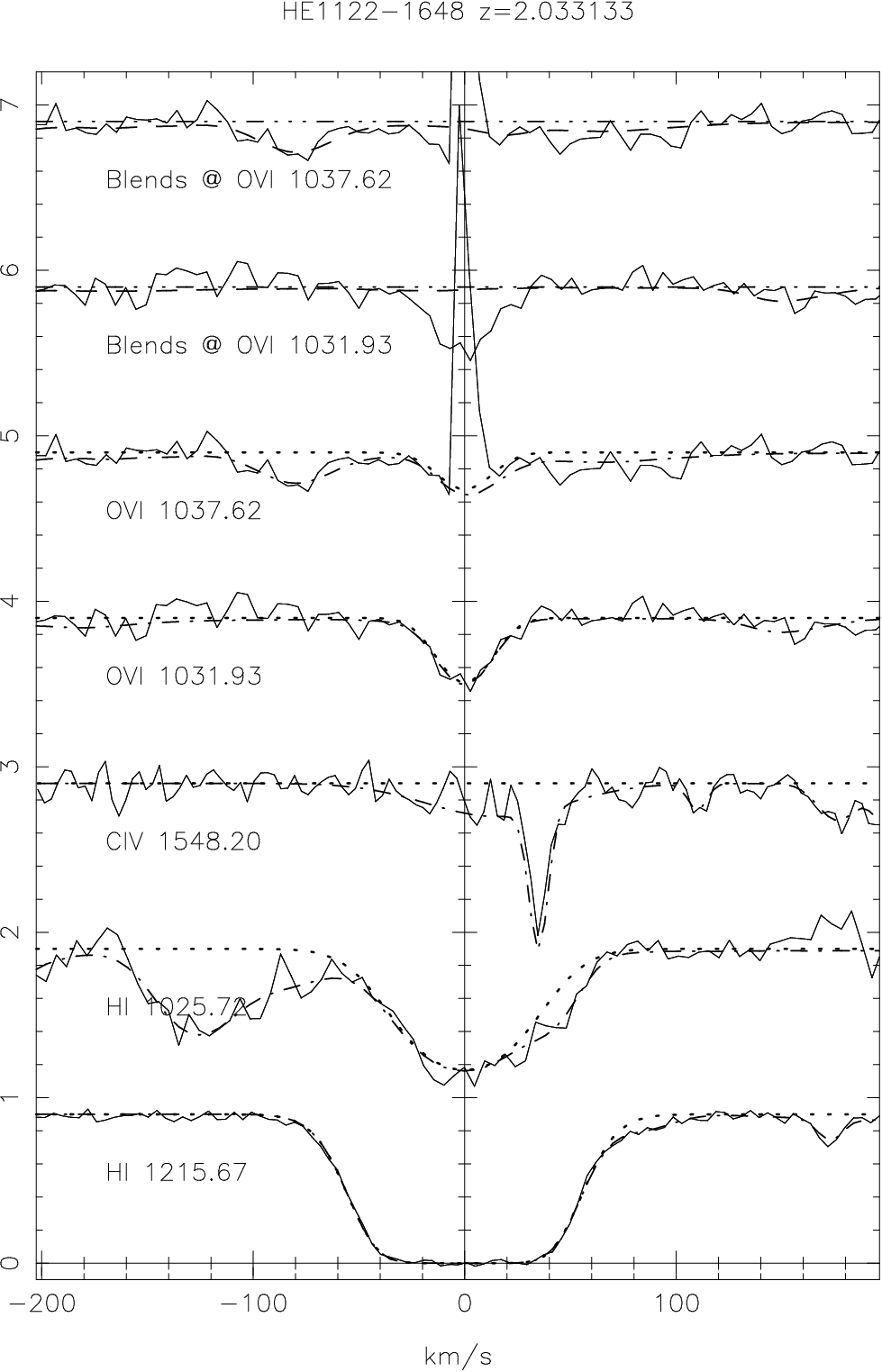}}}
\figcaption[z2p033p.ps]{ The system at $z=2.033133$ in HE1122--1648
showing the fitted \ion{H}{1} and \ion{O}{6} profiles. The \ion{C}{4} region
has been scaled by a factor five, so the zero level is 4.5 units below
the continuum. Note the noise spike at the position of \ion{O}{6}
1037.62. Other details are as for
Fig. \ref{fig:z2p074}. \label{fig:z2p033}}
\end{inlinefigure}

\begin{inlinefigure}
\centerline{\resizebox{0.75\colwidth}{!}{\includegraphics{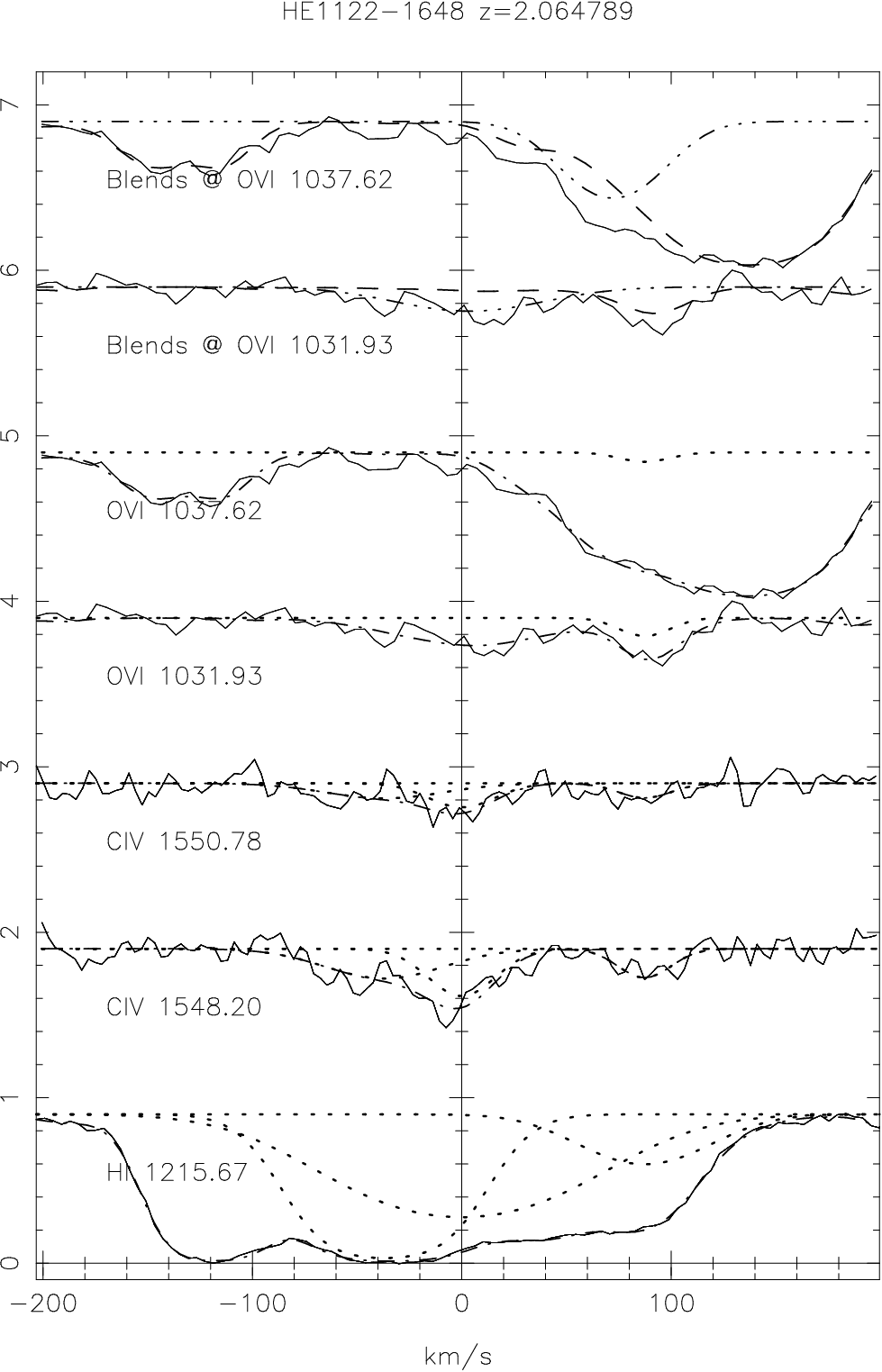}}}
\figcaption[z2p064p.ps]{ The systems at $z=2.064412$, 2.064789 and
2.065681 in HE1122--1648 \ion{H}{1}, \ion{C}{4} (scaled $\times 5$) and
\ion{O}{6} profiles. Details are as for Fig. \ref{fig:z2p074}.
\label{fig:z2p064}}
\end{inlinefigure}

\begin{inlinefigure}
\centerline{\resizebox{0.75\colwidth}{!}{\includegraphics{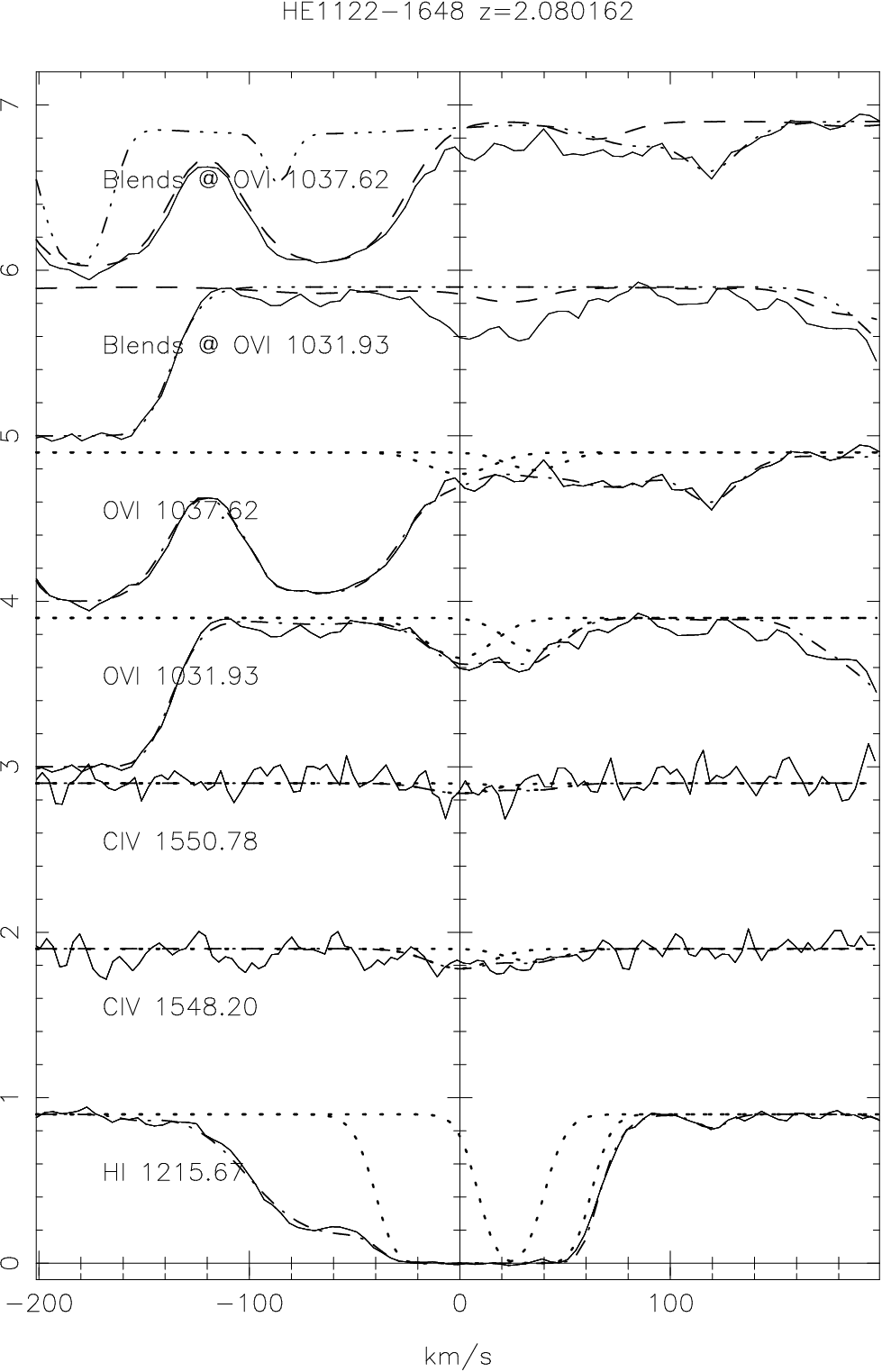}}}
\figcaption[z2p080p.ps]{ The systems at $z=2.080162$ and 2.080527 in
HE1122--1648 showing the fitted \ion{H}{1}, \ion{C}{4} (scaled $\times
5$) and \ion{O}{6} profiles. Other details are as for Fig.
\ref{fig:z2p074}.  \label{fig:z2p080}}
\end{inlinefigure}

\begin{inlinefigure}
\centerline{\resizebox{0.75\colwidth}{!}{\includegraphics{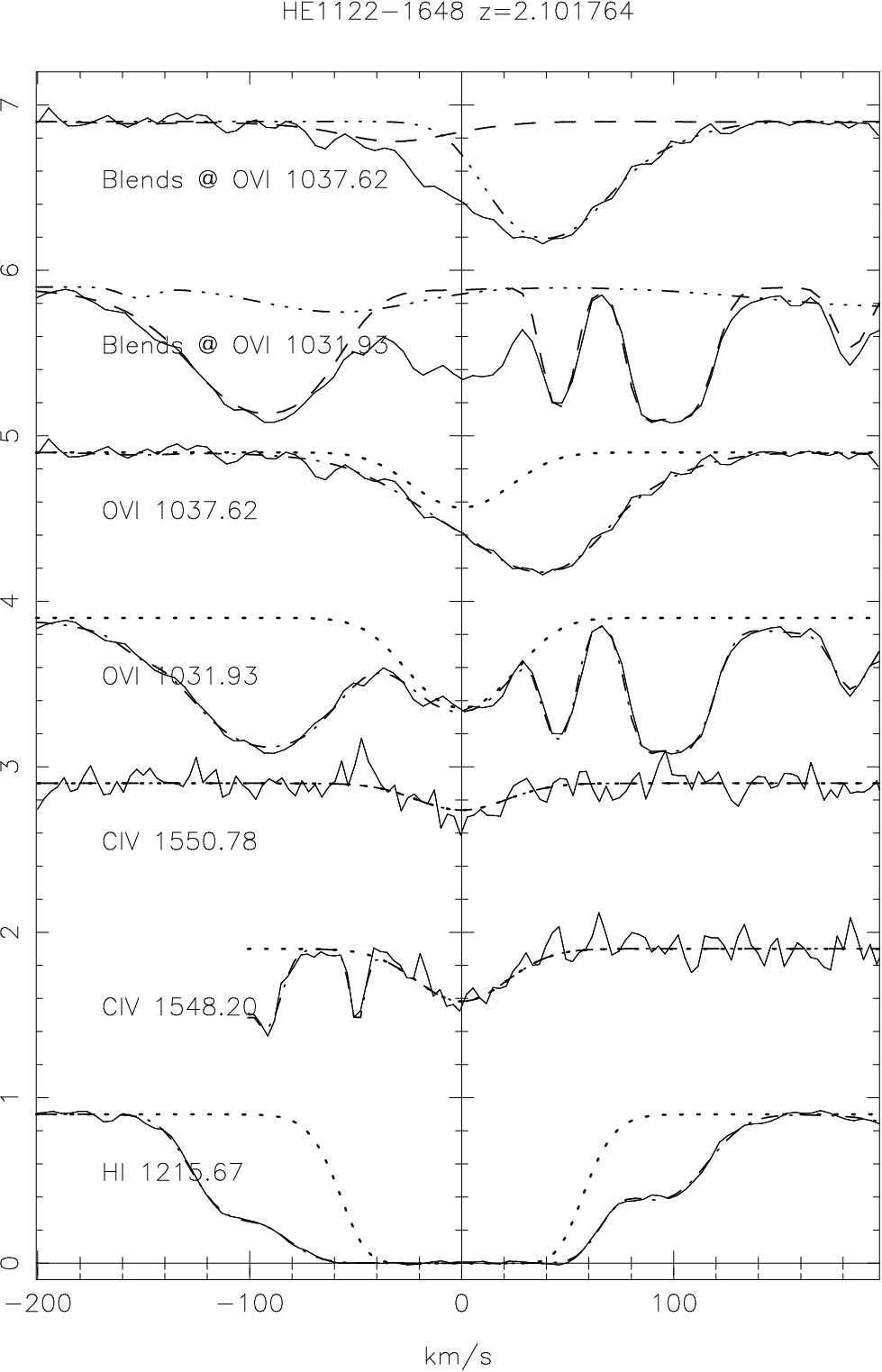}}}
\figcaption[z2p101p.ps]{ The system at $z=2.101764$ in HE1122--1648
showing the fitted \ion{H}{1}, \ion{C}{4} (scaled $\times 5$) and
\ion{O}{6} profiles. Details are as for Fig. \ref{fig:z2p074}.
\label{fig:z2p101}}
\end{inlinefigure}

\begin{inlinefigure}
\centerline{\resizebox{0.75\colwidth}{!}{\includegraphics{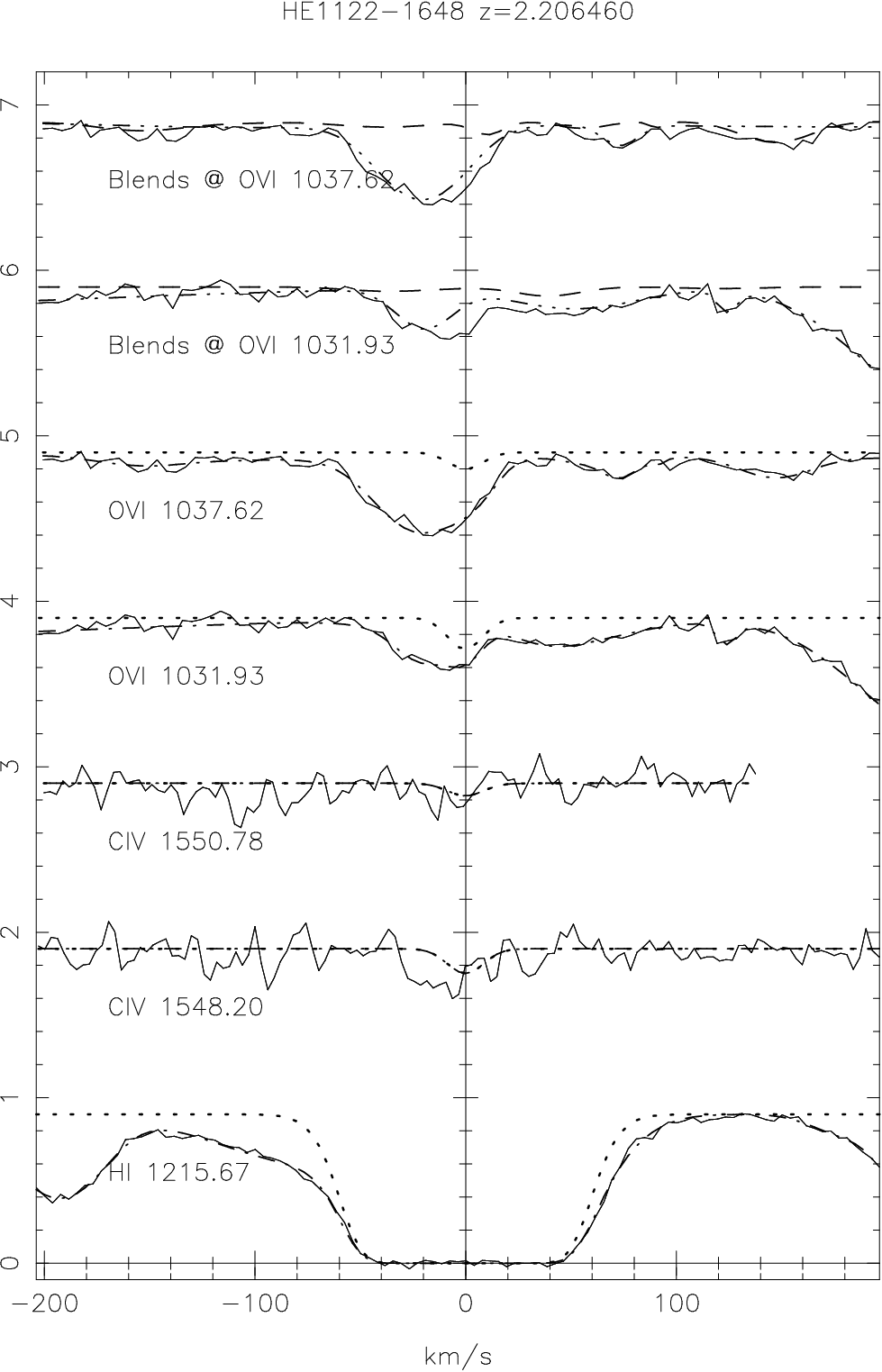}}}
\figcaption[z2p206p.ps]{ The system at $z=2.206460$ in HE1122--1648
showing the \ion{H}{1}, \ion{C}{4} (scaled $\times 5$) and
\ion{O}{6} profiles. Details are as for Fig \ref{fig:z2p074}.
\label{fig:z2p206}}
\end{inlinefigure}

\begin{inlinefigure}
\centerline{\resizebox{0.75\colwidth}{!}{\includegraphics{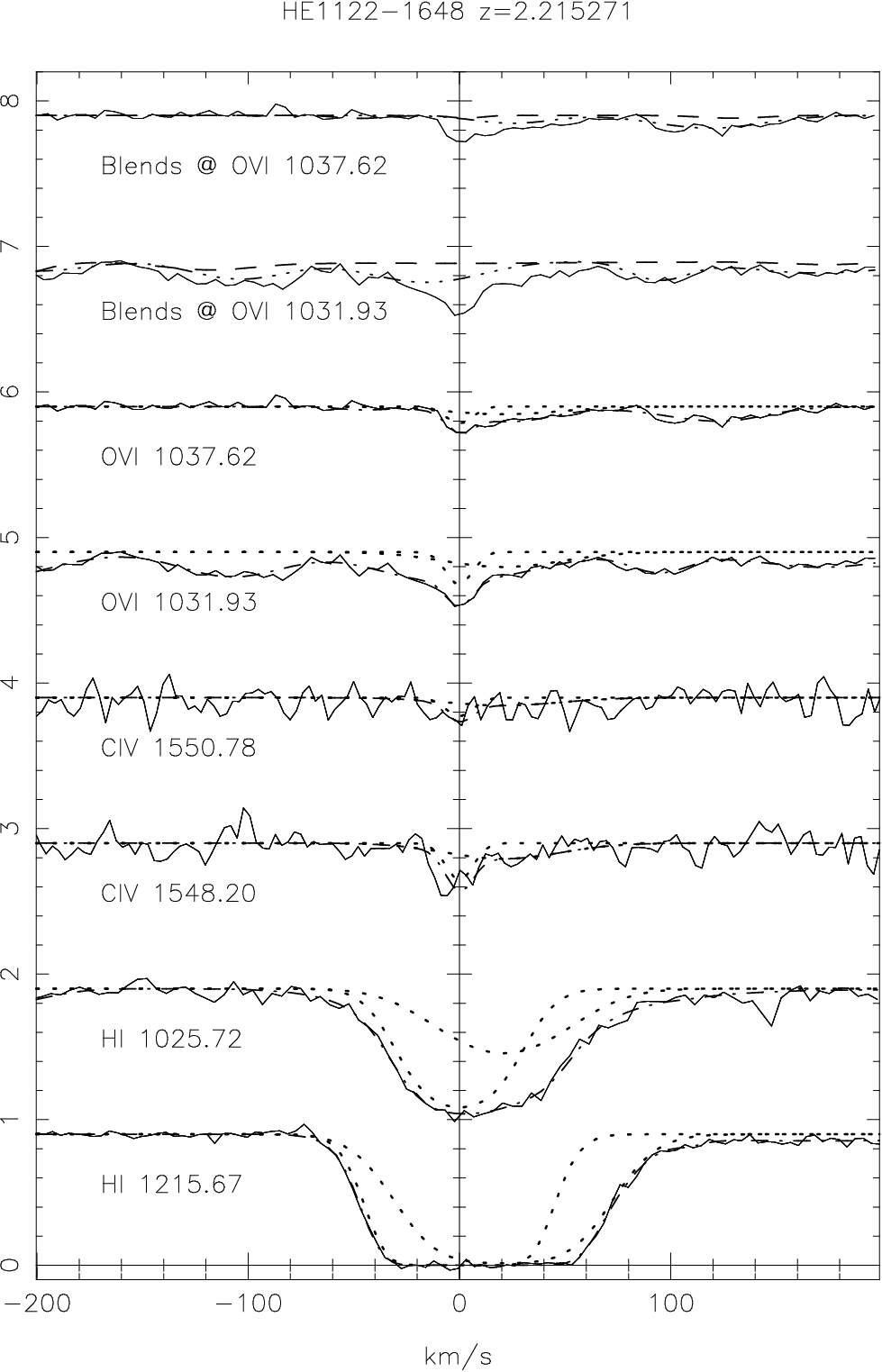}}}
\figcaption[z2p215p.ps]{
The systems at $z=2.215271$ and 2.215494 in HE1122--1648 showing the \ion{H}{1},
\ion{C}{4} (scaled $\times 8$) and \ion{O}{6} profiles. Details are as for Fig. \ref{fig:z2p074}. \label{fig:z2p215}}
\end{inlinefigure}

\begin{inlinefigure}
\centerline{\resizebox{0.75\colwidth}{!}{\includegraphics{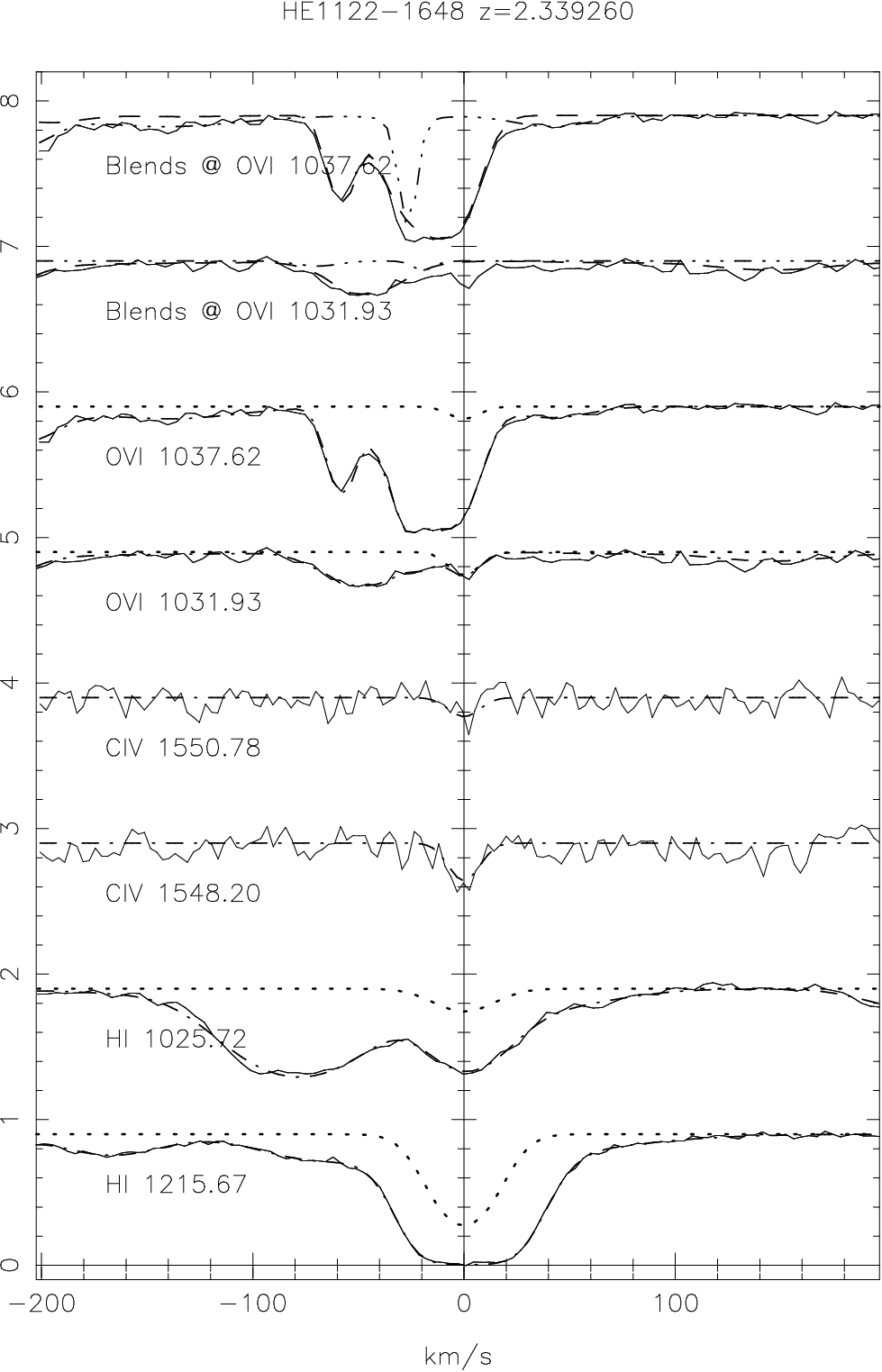}}}
\figcaption[z2p339p.ps]{
The system at $z=2.339257$ in HE1122--1648 showing the \ion{H}{1},
\ion{C}{4} (scaled $\times 8$) and \ion{O}{6} profiles. Details are as for Fig.
\ref{fig:z2p074}. \label{fig:z2p339}}
\end{inlinefigure}

\begin{inlinefigure}
\centerline{\resizebox{0.75\colwidth}{!}{\includegraphics{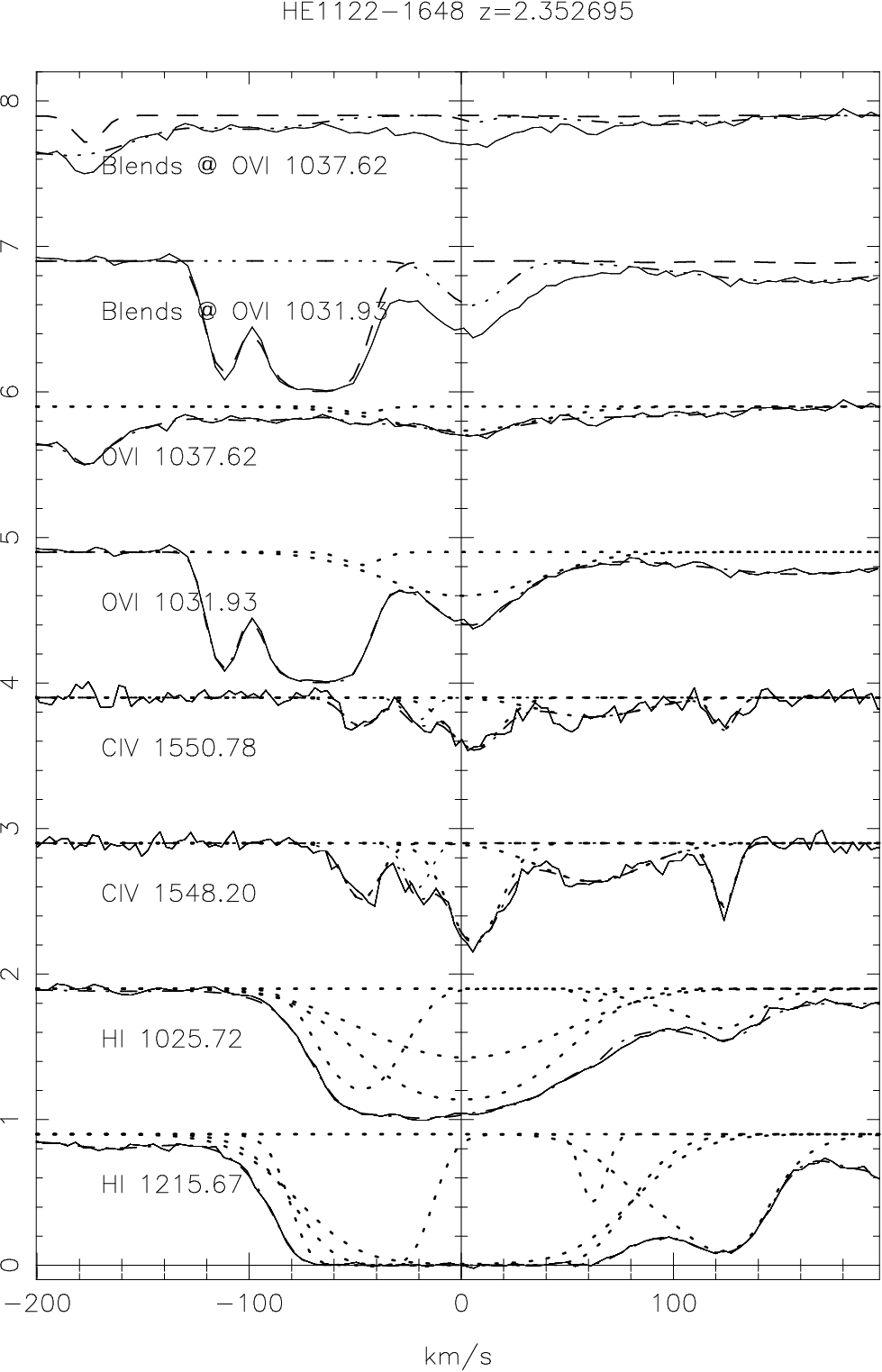}}}
\figcaption[z2p352.ps]{ The systems at $z=2.352170$ and 2.352695 in
HE1122--1648 showing the\ion{H}{1}, \ion{C}{4} (scaled $\times 5$) and
\ion{O}{6} profiles. Details are as for Fig. \ref{fig:z2p074}.
\label{fig:z2p352}}
\end{inlinefigure}

\section{Overall trends}

One of the features of the \ion{O}{6} search is that in nearly all
cases it is the high \ion{H}{1} column density systems which show these
lines.  If $\log N({\rm HI})\simgt 14.5$ then \ion{O}{6} is normally
clearly present. This may be the explanation for the most obvious
difference between the two sight lines; towards HE1122--1648 there are
10 distinct redshift systems with $z>2$ showing \ion{O}{6} in some
component, while towards HE2217--2818 there are only two.  The numbers
are too small for this difference to be statistically significant, but
it does correspond to the relative numbers of the high \ion{H}{1} column
density systems (14 and 6 respectively with $z>2$ and $\log N({\rm
HI})>14.5$).

\vspace{0.5cm}
\begin{inlinefigure}
\centerline{\resizebox{0.96\colwidth}{!}{\rotatebox{-90}{\includegraphics{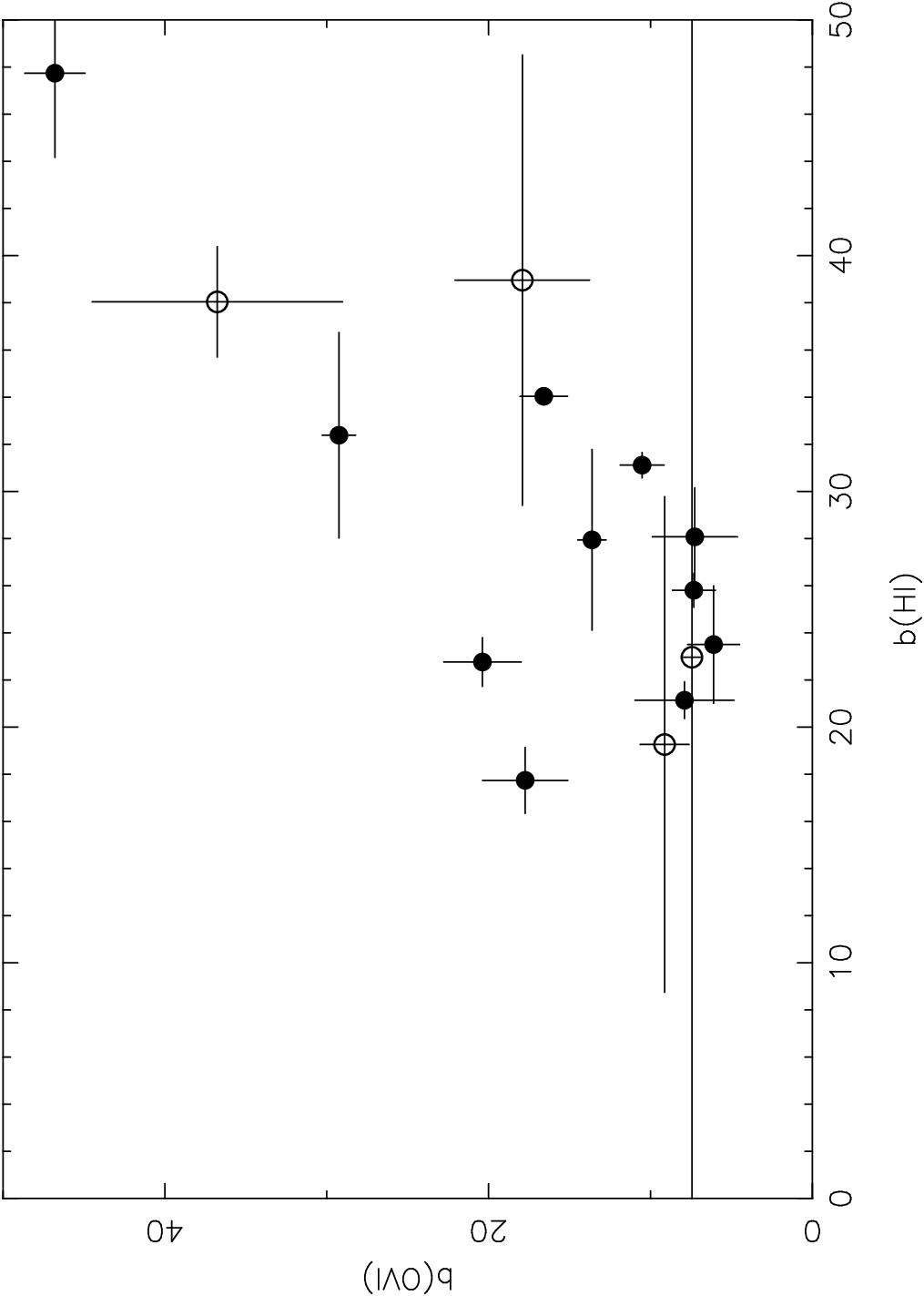}}}}
\figcaption[bOVIvbHI.ps]{The \ion{O}{6} Doppler parameter, $b$, in km~s$^{-1}$ vs the Doppler parameter for \ion{H}{1}. Other details are as for Fig. \ref{fig:NratvNHI}. \label{fig:bOVIvbHI}}
\end{inlinefigure}

The Doppler parameters for HI and OVI are correlated, as one would
expect from the way the data were analyzed. The turbulence dominated
systems have $b$(\ion{O}{6})$\sim b$(\ion{H}{1}), and for those where
thermal broadening dominates $b$(\ion{O}{6})$\sim {1\over 4}
b$(\ion{H}{1}). It appears from \ref{fig:bOVIvbHI} that both types of
system are present, and that either turbulence dominates strongly or
thermal broadening does. For thermal broadening at temperature of
$3\times 10^4$K $b$(\ion{H}{1})$= 22$~km~s$^{-1}$ and $b$(\ion{O}{6})$=
5.5$~km~s$^{-1}$. The group of systems with low \ion{O}{6}
Doppler parameters have temperatures close to this value.

Otherwise, it is hard to find any correlations at all. The
\ion{O}{6}/\ion{H}{1} column density ratio does appear to have an
inverse correlation  with the \ion{H}{1} column density (see Fig.
\ref{fig:NratvNHI}), so one might conclude that, as expected, the
ionization level is linked with the inferred density.  However, the
\ion{O}{6} systems have to be quite prominent to be detected deep in
the Lyman forest region of the spectrum and so most are near the
detection limit. The apparent correlation may therefore arise simply
because \OVI\ lines corresponding to absorption systems in the lower
part of the diagram are too weak to be detected.

\vspace{0.5cm}
\begin{inlinefigure}
\centerline{\resizebox{0.96\colwidth}{!}{\rotatebox{-90}{\includegraphics{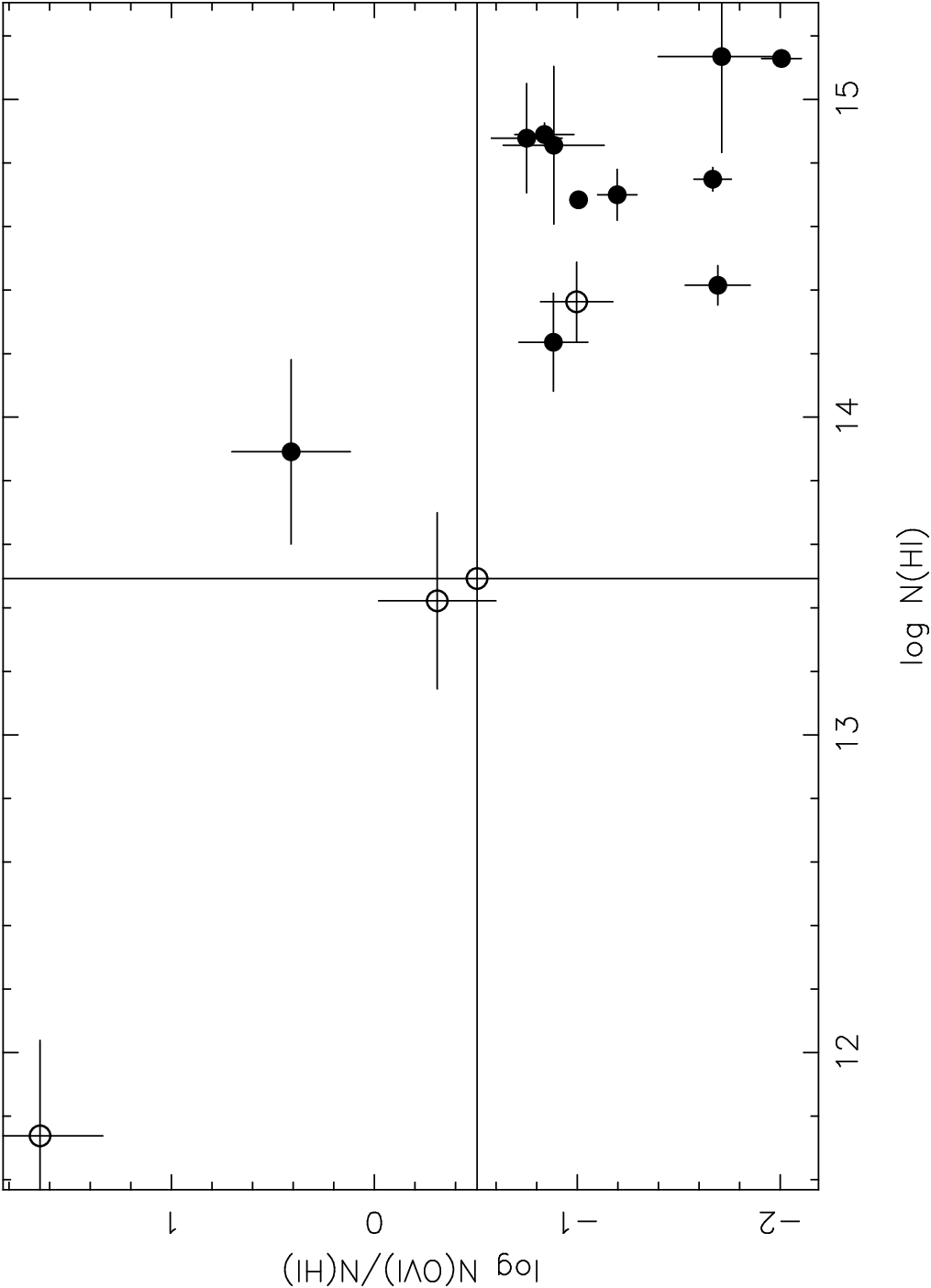}}}}
\figcaption[NratvNHI.ps]{The \ion{O}{6}/\ion{H}{1} column density ratio
vs that for \ion{H}{1} column density, with $1\sigma$ error estimates.
Where no error bars are shown, the error range is within the symbol
size. The open circles are for values where the \ion{O}{6}
identification is doubtful.  \label{fig:NratvNHI}}
\end{inlinefigure}

There are some correlations worth seeking since they could provide
clues to the nature of the systems. We caution that the number of data
points is small, so the absence of a significant correlation does not
mean that the underlying reasons for an expected trend are
incorrect.  One possibility is the \ion{O}{6} Doppler parameter against
the \ion{O}{6} column density (Fig.  \ref{fig:bOVIvNOVI}). Here a trend
could arise in several ways, such as undetected velocity structure in
higher column density systems, cooling rates linked to metallicity,
collisional ionization linking high ionization column densities and the
temperature, or just a selection effect since broad weak lines tend to
be lost in the noise. No correlation is seen.  Another is the inferred
temperature vs \ion{H}{1} column density, which might be expected if
the temperature depended strongly on density.  There is no such
correlation discernible with the available data (Fig.
\ref{fig:TvNHI}).\\
$\phantom{spacer}$\\
$\phantom{spacer}$\\
$\phantom{spacer}$\\

\vspace{0.5cm}
\begin{inlinefigure}
\centerline{\resizebox{0.96\colwidth}{!}{\rotatebox{-90}{\includegraphics{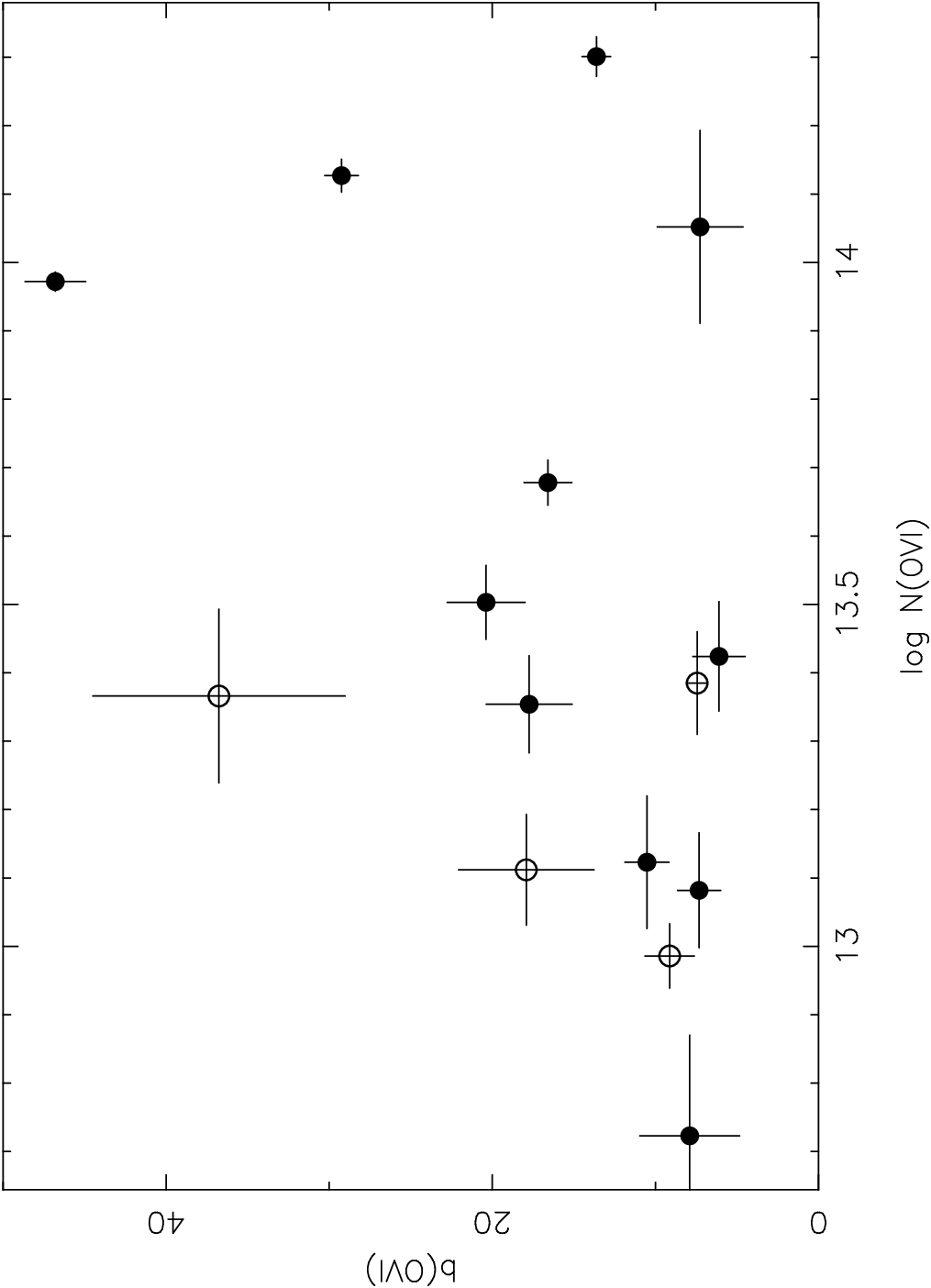}}}}
\figcaption[bOVIvNOVI.ps]{The \ion{O}{6} Doppler parameter, $b$, in km~s$^{-1}$ vs \ion{O}{6} column density. Other details are as for Fig. \ref{fig:NratvNHI}. \label{fig:bOVIvNOVI}}
\end{inlinefigure}

\vspace{0.5cm}
\begin{inlinefigure}
\centerline{\resizebox{0.96\colwidth}{!}{\rotatebox{-90}{\includegraphics{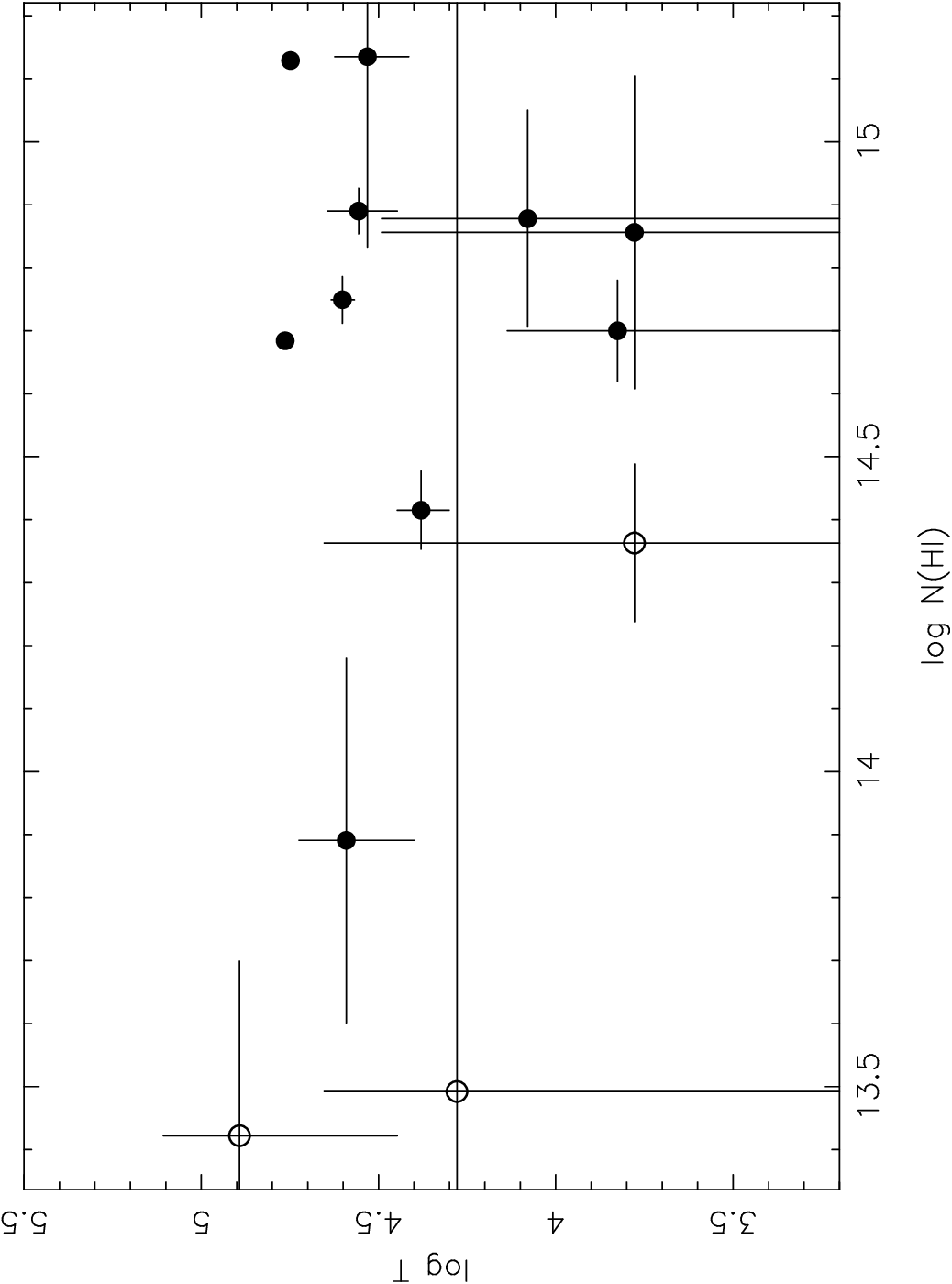}}}}
\figcaption[TvNHI.ps]{The inferred temperature vs \ion{H}{1} column density. Other details are as for Fig. \ref{fig:NratvNHI}. \label{fig:TvNHI}}
\end{inlinefigure}

\section{Modeling the \ion{O}{6} systems}

We have run
CLOUDY\footnote{\texttt{http://www.pa.uky.edu/$\sim$gary/cloudy/}}
(version 94.00) \cite{Ferland99} models for various forms of the
ionizing flux using four example systems from HE$1122-1648$ where the
\ion{C}{4} and \ion{O}{6} appear to have the same velocity structure.
These were chosen to cover a range of \ion{H}{1} column densities and
\ion{O}{6}/\ion{H}{1} ratios and they are not necessarily the ones with
the most reliable or accurately determined parameters.  The details are
given in Table \ref{tab:OVImod}.  In this table the row
containing the redshift identifies the system being considered, and
gives the measured parameters from Table \ref{tab:OVI1122}, along with
a density estimate based on the \ion{H}{1} column density using a
prescription from Schaye (2001), as detailed below.

Subsequent lines in the table give the results of models for which the
heavy element abundances and hydrogen density were the model parameters
varied to give the best fit to the observed \ion{H}{1}, \ion{C}{4}
(where measured) and \ion{O}{6} column densities. The models were
chosen to have all heavy elements present with solar relative
abundances, but the heavy element/hydrogen ratio was allowed to vary to
obtain a best fit to the data, using the OPTIMIZE command in CLOUDY.
The log of the best fit C/H value relative to solar is given in the
[C/H] column in the table. The other parameter which was allowed to
vary to obtain a best fit is the hydrogen number density, $n_{\rm H}$.  Since
there are at most two independent ion ratios constraining the model
parameters, and two model parameters, a satisfactory fit is always
possible. If \ion{C}{4} is undetected or unmeasurable then there are
insufficient constraints to allow an estimation of both the metallicity
and the hydrogen density.

The ionizing backgrounds chosen for these calculations are the
Haardt-Madau model for the ultraviolet background radiation at $z=2$
from quasars \cite{Haardt96}, and simplified power law models. We have
generally chosen a power law slope of $-1.8$ \cite{Zheng97} as the best
estimate for a quasar background ionizing spectrum, and in some cases a
power law slope of $-1.5$ has been used to see how sensitive the
results are to this choice. The effects of \ion{He}{2} absorption are
dealt with in a simple way by inserting a break so that the flux just
above 4 Rydbergs is $10^{-B}\times$ the flux just below, with the same
power law on either side of the break.  For $B$ we chose the values 0
(no break), 1, and 2.  Note that for a spectral index $\alpha$, a break
$B$ corresponds to a softness parameter $S \equiv \Gamma_{\rm HI} /
\Gamma_{\rm HeII} \approx 4^{\alpha + 1} 10^B$, where $\Gamma_{\rm HI}$
and $\Gamma_{\rm HeII}$ are the ionization rates of \ion{H}{1} and
\ion{He}{2} respectively.  The $z=2$ Haardt-Madau model has a softness
parameter $S\approx 92$, intermediate between the $B=0$ and $B=1$ power
law models.  The model designation in the table gives the power law and
break strength. For example, model ``P1.8B1'' has a power law slope of
$-1.8$, and a break of 1 (i.e., a factor 10 in flux) at the \ion{He}{2}
Lyman limit.  If the break is large (3, and occasionally 2 for the
steeper power laws) there is very little high energy ionizing radiation
and \ion{O}{6} is effectively suppressed. In those cases the \ion{O}{6}
column density is close to zero and a fit to the data cannot be
obtained within the operational range of the CLOUDY program.  Such
cases are omitted from the tables. 
In \S\ref{sec:densities} we will show that photoionization by spectra
with $S\gg 10^2$ results in absorber sizes that are inconsistent
with the data.

The flux normalization was chosen to
be $\log F_{\nu}=-20.31$ (erg~cm$^{-2}$~s$^{-1}$~Hz$^{-1}$) at 1
Rydberg in all power law cases. This corresponds to $\log
J_{\nu}=-21.41$ (erg~cm$^{-2}$~s$^{-1}$~Hz$^{-1}$~sr$^{-1}$)
and the photoionization
rate for neutral hydrogen, $\log\Gamma_{\rm HI}\approx -12.02$, matches
the Haardt-Madau value if the power law slope is $-1.8$. 
If a different $F_{\nu}$
value is preferred, then the results in the tables still apply, now for
hydrogen densities rescaled so that the ionization parameter remains
constant.

We may transform the \ion{H}{1} column densities to densities if we assume
photoionization and (local) hydrostatic equilibrium \cite{Schaye01}.
This transformation is a function of the ionizing background, but we
can use it here to estimate the background if we believe the models, or
test the models if we believe the background, or use it to estimate the
oxygen/carbon ratio compared with the solar value. 
For temperatures of $\sim 5\times
10^4$~K, and $\Gamma_{\rm HI}\sim 10^{-12.02}$, and choosing his default
value for the fraction of the mass in gas ($f_g = 0.16$), Schaye's
equation (8) linking the \ion{H}{1} column density $N_{\rm HI}$ with the
hydrogen density $n_{\rm H}$ becomes
\begin{equation}
\log n_{\rm H}\sim {2\over 3}\log N_{\rm HI} - 13.8.
\label{eq:nH-NHI}
\end{equation}
We have used these density estimates to provide ``observed'' densities
in tables \ref{tab:OVImod} and \ref{tab:OVImmx} for
comparison with those given by the best fit models to the ion ratios.
Assuming $\Omega_bh^2=0.02$, the mean hydrogen number density at $z=2$
is $\sim 5\times 10^{-6}~{\rm cm}^{-3}$, with a corresponding $N_{\rm
HI}\sim 6\times 10^{12}$, so in all the cases considered here the
regions have a significant over-density.

\subsection{Constraints based on \ion{H}{1}, \ion{C}{4} and \ion{O}{6}}
 
For a given background ionizing flux the errors in the hydrogen density
and heavy element abundances given in Table \ref{tab:OVImod} depend on
the precision with which the column densities are determined. Given the
assumption that the relative heavy element abundances are solar,
changing the heavy ion/\ion{H}{1} ratio effectively determines the
heavy element abundances, while the \ion{C}{4}/\ion{O}{6} ratio
determines the ionization parameter and so the hydrogen density.  For
example, for the $z=2.101764$ system in HE1122--1648 the error estimate
of $0.17$ in $\log N_{\rm HI}$ becomes an uncertainty of about $0.17$
in the heavy element abundances. The errors of $0.02$ in $\log N_{\rm
CIV}$ and $\log N_{\rm OVI}$ translate to an error of about $0.03$ in
$\log n_{\rm H}$.  Thus, for a given background flux the hydrogen
density can be quite precisely determined with the model assumptions we
have made.  However, the formal errors in the column densities and
other quantities given in Tables \ref{tab:OVI2217} and
\ref{tab:OVI1122} are based on the assumption that the Voigt component
model is correct in all respects, including in particular the number of
\ion{H}{1} components blended with the \ion{O}{6} lines. It is quite
possible in the $z=2.101764$ case that the \ion{O}{6} column density is
lower if there is additional \ion{H}{1} blended with the \ion{O}{6}
lines. If we choose, for example, to reduce the \ion{O}{6} column
density arbitrarily by a factor of two so $\log N_{\rm OVI}$ becomes
13.83 without changing the \ion{C}{4} column density, then, for the
Haardt-Madau background flux, the the best fit hydrogen density becomes
$\log n_{\rm H}=-4.00$ and [C/H]$=-1.99$.

Collisional ionization models were also attempted, and the results for
these are given the model designation ``Coll'' in Table
\ref{tab:OVImod}. The value in the temperature column is then the best
fit temperature for the model. In no cases are the temperatures
inferred from the line widths high enough to produce any significant
collisionally ionized \ion{O}{6}.

A further possibility is that different absorption lines arise in
different parts of the cloud model, so our assumption of velocity
agreement may not be valid. For the CLOUDY models we have checked this
point by examining the ion fractions through the cloud model, and find
that the ionization fractions of the species we are modeling,
\ion{H}{1}, \ion{O}{6} and \ion{C}{4}, vary by at most 10\% through the
cloud.  This does, however, not preclude the possibility that different
ionization levels come from different regions in a system with
substructure.  This point needs further investigation, but the fact
that there is general agreement between the ion redshifts when fitted
independently suggests this is not a very important concern.

There is also no guarantee that the C/O ratio in these systems is close
to the solar value of $\log({\rm C/O})=-0.37$, as assumed for the
models. Since the enrichment is likely to have occurred from massive
stars, the C/O ratio may well be smaller than the solar value. With
examples of oxygen-rich supernova remnants as a guide (see e.g.\ Blair
et al., 2000) suggesting that the C/O ratio can be about an order of
magnitude smaller, we have run power law cases with oxygen enhanced so
that $\log({\rm C/O})=-1.37$ while all other heavy elements have solar
abundance ratios with respect to carbon. These are the models labeled
``OP....'' in Table \ref{tab:OVImod}, and the number in the ``metals''
column is then [C/H]. The net effect is an increase in the hydrogen
density to give the change in ionization then required to reproduce the
observed \ion{O}{6}/\ion{C}{4} ratio.

\subsection{Temperature comparisons}

We now have temperature and density parameters inferred from the data
for each system, and the same quantities, together with the element
abundances, are predicted by the models.  Before exploring the
agreement, or otherwise, between the two, it is worth examining the
uncertainties in the two quantities inferred from the data in a little
more detail.

An important first question is: could the \ion{O}{6} arise from
collisional ionization? As can be seen from a comparison of the
temperatures inferred from the Voigt profile fits (Tables
\ref{tab:OVI2217} and \ref{tab:OVI1122}) and those needed for
collisionally ionized models (Table \ref{tab:OVImod}), the temperatures
inferred from the line widths are generally much too low to allow this
as a possibility. However, the formal errors in the temperature
estimates are based on the assumption that the decomposition into
velocity components as separate clouds, each with a single temperature,
is correct, and we know from models of the evolution of structure that
this is an oversimplified picture. And even if it were not, the
decomposition into Voigt profile components is not unique.  Since the
components are optically thin the CLOUDY models based on the the
relative column densities will still be broadly applicable, provided
there are no large relative abundance variations between the components
of what we treat as a single entity. However, the temperatures based
directly on the line widths could be seriously affected. This is
particularly true where there is a large turbulent component in the fit
to the line widths, or when the lines are significantly blended.

Temperatures $T\sim 2.3$-$2.5\times 10^5$~K are required if collisional
ionization dominates, so these temperatures were imposed on each
system. Acceptable Voigt profile fits to the \ion{O}{6} lines were
found in all four of the test cases in Table \ref{tab:OVImod}, either
because the Doppler parameters for the heavy elements are in any case
large enough to accommodate such temperatures or, in the exceptional
case of the $z=2.206460$ system in HE1122--1648, because other lines
blended with the \ion{O}{6} adjusted to accommodate the constraint and
\ion{C}{4} is too weak to provide much additional information. The
results are given in Table \ref{tab:OVIhiT}. Where the column density
value is blank the error column gives the log of the $1\sigma$ upper
limit.

The common feature to all of these high temperature models is that the
Lyman lines associated with the \ion{O}{6} are too broad to be
significant contributors to the observed \ion{H}{1} line profiles.
Additional, cooler \ion{H}{1} components which contain the bulk of the
\ion{H}{1} column density are invariably required because the
\ion{H}{1} column density associated with the \ion{C}{4} and \ion{O}{6}
must be low.  Strictly speaking this new system would require
remodeling to estimate abundances and collisional temperature, and then
we should redo the fit and so on until the process converges, but we
have not done this. The temperature is effectively set by the
\ion{C}{4}/\ion{O}{6} ratio, and this can change somewhat when the
\ion{H}{1} constraint on redshift and Doppler parameter is changed.
However, the observed range is satisfied by a narrow range of
temperatures $T\sim 2.2$-$2.5\times 10^5$~K, so the predicted line
widths will not be very different.  The reduced \ion{H}{1} column
density then requires a model with higher heavy element abundances in
the collisionally ionized component.  If we require that the \ion{O}{6}
is collisionally ionized, then each system showing \ion{O}{6} must have
associated with it a cooler system with higher \ion{H}{1} column
density and no detected heavy elements. Note also that in some of the
cases given in Table \ref{tab:OVI1122} the \ion{C}{4} and \ion{O}{6}
Doppler parameters are significantly less than 15 km~s$^{-1}$, and so
do not allow the high temperatures required.

At first sight even some of the photoionized models have temperatures
which are too high. For example, for the $z=2.101764$ system the
best-fit temperature is $1.3\pm 1.9\times 10^4$ km~s$^{-1}$, while the
modeled temperatures are in excess of $4\times 10^4$~K, which appears to
rule out those models at roughly the $2\sigma$ level. However a look at
the independently determined Doppler parameters for each ion in that
system gives a hint to what may be happening. They are all much larger
than the thermal Doppler parameters, and so bulk motion dominates the
line widths. The best-fit turbulent component has a Doppler width of
$29.1\pm 1.2$ km~s$^{-1}$, while even the \ion{H}{1} thermal Doppler
parameter is $\sim 15$ km~s$^{-1}$. If we demand that the temperature
for this \ion{O}{6} absorber is 50,000~K, for example, then a
satisfactory fit to the data is obtained by introducing two velocity
components at this temperature. Therefore, the temperature constraint
is not one which is a significant concern for this system.

The pair of systems at $z=2.080162$ and $2.080527$ cannot be dealt with
in this way since the \ion{H}{1} lines are themselves too narrow to
accommodate two components at a much higher temperature than about
20,000~K. It is possible to add a third, dominant \ion{H}{1} component
without associated heavy elements and obtain a satisfactory fit.
However, the CLOUDY photoionization model equilibrium temperatures may
not be the ones we should be attempting to match. In the hydrodynamic
simulations adiabatic cooling as the universe expands is still
important at these column densities. This cooling process is not
included in the CLOUDY models, so temperatures they give are too high.

\subsection{Density comparisons}
\label{sec:densities}

If the absorbers are photoionized and in approximate, (local)
hydrostatic equilibrium, then the densities predicted by the CLOUDY
models should roughly agree with those predicted by the density/\ion{H}{1}
column density relation [equation (\ref{eq:nH-NHI})].  From Table
\ref{tab:OVImod} we can see that generally the models with harder far
ultraviolet flux are preferred, i.e., those with no or a small change
in intensity at 4 Rydbergs, and those with a flatter power law
spectrum. The Haardt-Madau (1996) background spectrum produces models
with densities to within about a factor two of the Schaye (2001)
estimate. The pure power law models with no break and slope $-1.8$ also
give density estimates in reasonable agreement, and for power law
models with a slope $-1.5$ a break at the \ion{He}{2} edge of a factor less
than 10 is preferred.  Any strong break in a power law spectrum gives
densities that are significantly lower than expected.

We can place a lower limit on the density from the bulk motion
component, $b_{\rm turb}$, of the line widths. The velocity width across an absorber of characteristic size $\ell$ due to the Hubble flow is
\begin{equation}
b_{\cal H}\sim H(z) \ell/2
= H(z) N_{\rm H} /2n_{\rm H} 
= H(z) N_{\rm HI}/2n_{\rm HI}. 
\label{eq:bHub}
\end{equation}
Using equation 6 from Schaye (2001) for the neutral/total hydrogen
fraction of a highly ionized, optically thin plasma gives
\begin{equation}
b_{\cal H}\sim 11\,h ~ {\rm km}\,{\rm s}^{-1} \left({N_{\rm HI}\over
10^{15}~{\rm cm^{-2}}}\right)\left({n_{\rm H}\over 10^{-4}~{\rm cm^{-3}}}\right)^{-2}T_4^{0.76}\Gamma_{12},
\end{equation}
where $T_4$ is the temperature in units of $10^4$~K, $\Gamma_{12}$ is the 
\ion{H}{1} ionization rate in units of $10^{-12}$~s$^{-1}$, and we assumed
$(z,\Omega_m,\Omega_\Lambda) = (2.2,0.3,0.7)$.

For systems with $N_{\rm HI}\sim 10^{15}$~cm$^{-2}$ and densities less
than $\sim 10^{-4.5}$~cm$^{-3}$, the velocities involved are $\sim
100~{\rm km}\,{\rm s}^{-1}$ or more. Peculiar velocities can change
this somewhat, but the Hubble velocity width is so sensitive to
density that the upper limit $b_{\rm turb}\sim 40~{\rm km}\,{\rm
s}^{-1}$ (Tables \ref{tab:OVI2217} and \ref{tab:OVI1122}) effectively
precludes the soft spectra models (..B1 and ..B2) given in Table
\ref{tab:OVImod}.

Note that both the Schaye (2001) and CLOUDY model densities have the
same dependence on the normalization of the ionizing flux, so the
agreement between the two does not constrain this quantity in any way.
The comparison is, as we have seen, sensitive to the shape of the
ionizing background. So what we have shown is that, to obtain a
consistent picture, the ionizing background at $z\sim 2$ could have,
from the choices we have explored, the Haardt-Madau form or a power law
with a small jump at the \ion{He}{2} edge.

The density estimates of both the photoionization models and Schaye (2001)
assume that the absorbers are illuminated only by the mean background
radiation. If local sources of UV photons are significant, then the
density will be underestimated. The $2.007$ system in HE$1122-1648$
illustrates the possible range well. The \ion{O}{6} is in a noisy
region, so it was not selected for detailed analysis, but the presence
of \ion{O}{6} in one system and \ion{Si}{4} in another with comparable
\ion{H}{1} column density suggests that the densities in the two
components must be very different if the ionization source is the same
for both. Table \ref{tab:OVImmx} shows that for power law ionization
the density difference is at least $1.5$ orders of magnitude and large
abundance differences are also required.  Attempts to fit this complex
as a single cloud fail. Unless there are large abundance anomalies it
is not possible for \ion{O}{6} and \ion{Si}{4} to coexist in a cloud
with \ion{H}{1} column density $\log N({\rm HI})\sim 15.5$.

\subsection{The high ionization system at $z=2.030096$ in HE$1122-1648$}

The system at $z=2.030096$ system in HE$1122-1648$ shows \ion{C}{4},
\ion{N}{5}, and possibly \ion{O}{6} absorption, while Ly$\alpha$ is
very weak ($\log N({\rm HI}) = 11.7 \pm 0.3$). This contrasts with
most of the reliable systems we have 
found, where Ly$\alpha$ is strong and \ion{N}{5} is not detected. The
line widths are such that a high temperature is possible, and so
collisional ionization could be important. 

The density inferred from the photoionization models ($\log n_{\rm H} =
-4.11$ for model HM96, see Table \ref{tab:OVImodX}) is higher than one
would expect for a photoionized, gravitationally confined cloud with
the low \ion{H}{1} column density seen. The neutral hydrogen fraction
$\log(n_{\rm HI}/n_{\rm H}) =-4.59$, so with this density the cloud is
very small:  $\sim 10^2$~ pc, and has a characteristic mass of only
$\sim$1~M$_\odot$.  Using equation (8) of Schaye (2001), we see that
for the cloud to be self-gravitating, the gas fraction would have to be
extremely small, $f_g \sim 10^{-7} (n_H / 10^{-4.11}~{\rm
cm}^{-3})^{-3} \Gamma_{12}^2 T_4^{0.52}$, and the total mass would thus
be of order $10^7$~M$_\odot$.  If the gas fraction is significantly
greater than this, then an external pressure of $P/k \sim 1~{\rm
cm}^{-3}\,\K$ would be required to confine the gas. This system may
thus be a high redshift analogue of the low redshift system of Tripp et
al.\ (2002) or the weak \MgII\ systems studied by Rigby, Charlton, \&
Churchill (2002).

Since the ionization balance of a collisionally ionized plasma depends
only on the temperature, we cannot constrain the density if the system
is collisionally ionized. Nevertheless, we can check whether the cloud
can be self-gravitating by using the constraint that the differential Hubble
flow across the absorber must be smaller than the observed line width.
Combining equation (\ref{eq:bHub}) with equation (4) of Schaye (2001),
which gives $n_H(N_H,T,f_g)$ for a self-gravitating cloud, yields the
Hubble flow velocity across the system, 
\begin{equation} 
b_{\cal H}\sim 1.3\, h ~{\rm km}\,{\rm s}^{-1}~\left({N_{\rm H}\over
10^{20}~{\rm cm}^{-2}}\right)^{-1} T_4 \left({f_g\over 0.16}\right).
\end{equation}
For collisional
ionization at a temperature $T=2.35\times 10^5$~K the neutral hydrogen
fraction is $10^{-5.7}$, so for this system $N_{\rm H}=10^{17.4}~{\rm
cm}^{-2}$. Then 
$b_{\cal H}\sim 1.2\times 10^4 h ~{\rm km}\,{\rm
s}^{-1}(f_g/0.16)$. 
Thus, such a collisionally ionized system can only be self-gravitating
if the gas fraction is extremely small. Note that if $b_{\cal H}$ is
much greater than the sound speed ($\sim 80~{\rm km}\,{\rm s}^{-1}$ in
this case), the absorber size exceeds the sound horizon, thermal
pressure is irrelevant and the cloud cannot be in local,
hydrostatic equilibrium (see Schaye 2001).

Both collisional and
photoionization models (see Table \ref{tab:OVImodX}) require that the
heavy element abundances be high -- somewhere in the range 0.5 to 5
$\times$ solar. This system is also interesting in that the nitrogen
abundance is high, and may even be enhanced relative to that of carbon
and oxygen. 

In summary, regardless of whether the system is photoionized or
collisionally ionized, its metallicity is high and it can only be
self-gravitating if it has a negligible gas fraction. Alternatively, it
could be confined by external pressure or it could be a freely
expanding transient.

\section{Cosmological Implications}

An estimate the redshift evolution of the number of \ion{O}{6} systems
between $z\sim 0$ and $z\sim 2$ may be made by using the Tripp et al.
(2000) low redshift results in conjunction with the data presented
here.  Tripp et al. find that the number per unit redshift ${d{\cal
N}\over dz}\sim 48$ for those systems in which the rest equivalent
width $W_r>30$~mA for both OVI lines in the redshift range $0.17\leq
z\leq 0.27$. With 90\% confidence they set ${d{\cal N}\over dz}\geq
17$. The systems with redshifts $z=2.101764$ and 2.352695 in
HE1122-1648, and 2.074740 and 2.075446 in HE2217-2818 satisfy their
equivalent width criterion. This gives ${d{\cal N}\over dz}= 5.7 \pm
2.9$, which is considerably lower than Tripp et al's value.  The
spectral resolution is comparable, but the S/N in the Tripp et al. data
is not as high, so some of the blends separated in tables
\ref{tab:OVI2217} and \ref{tab:OVI1122} might have been detected as
single systems.  Likely candidates are the close pairs at $2.080$ and
at $z=2.214 5$ in HE1122-1648, and  the complex at z=2.180-2.182 in
2217 would provide two systems above the threshold, as blends of the
lowest two and highest two redshifted OVI's in the complex. Then there
are a possible 8 \ion{O}{6} systems above the threshold, so ${d{\cal
N}\over dz}= 11.4\pm 4.0$. This is consistent with the number density
of such systems found in the redshift range $1.21\leq z\leq 1.67$ by
Reimers et al. (2001), and strengthens the suggestion that the number
density of \ion{O}{6} systems decreases with increasing redshift.
However, there are large uncertainties in the number densities in the
samples so further investigation is needed.

It is interesting to check whether the identified \ion{O}{6} systems contain a
cosmologically significant fraction of metals and/or baryons.
The \ion{O}{6} density relative to the critical density is given by
\begin{eqnarray}
\Omega_{\rm OVI}&=&{8\pi G m_{\rm O}\over 3 H_0c}\int{ {d^2 n\over dN_{\rm OVI}dz}{H(z)\over H_0(1+z)^2}dN_{\rm OVI}}\nonumber\\
&\approx&{8\pi G m_{\rm O}\over 3 H_0^2c} {H(z)\over (1+z)^2} {\Sigma N_{\rm OVI}\over\Delta z}.\nonumber
\end{eqnarray}
For two redshift 
ranges centered on $z=2.2$, $\Delta z=0.365$ and 0.360,
we have a mean total \ion{O}{6} column density per unit redshift $\Sigma
N_{\rm OVI}/\Delta z = 1.26\times 10^{15}$, and so $\Omega_{\rm OVI}=
6\times 10^{-8}/h$.
For the photoionization models $\log{\rm O\over OVI}\sim 0.75$, so the
cosmic metallicity at $z=2.2$ due to \ion{O}{6} systems is 
\begin{eqnarray}
[{\rm O/H}]&=&{\Omega_{\rm OVI}\over\Omega_b}{N_{\rm O}\over N_{\rm
OVI}}{m_{\rm H}\over m_{\rm O}(1-Y)}{1\over ({\rm
O/H})_{\odot}}, \\ 
&\sim&1.6\times 10^{-3}h\nonumber, 
\end{eqnarray}
where we used $\Omega_m=0.3$, $\Omega_{\Lambda}=0.7$, and $\Omega_bh^2
= 0.02$. 
The baryon density associated with the detected \ion{O}{6} systems is
\begin{equation}
\Omega_{\rm g,OVI}\approx{8\pi G m_{\rm H}\over 3 H_0^2c(1-Y)} {n_{\rm
H}\over n_{\rm HI}} 
{H(z)\over (1+z)^2} {\Sigma N_{\rm HI}\over\Delta z}.
\end{equation}

The CLOUDY models give the total hydrogen/\ion{H}{1} ratio $n_{\rm H}/n_{\rm
HI}\sim 10^5$.  We do not have \ion{H}{1} measurements directly associated
with the \ion{O}{6} component for two of the systems in HE$2217-2818$,
so we base the estimate on the systems towards HE$1122-1648$.  Then
over a redshift range $\Delta z=0.365$ the total \ion{H}{1} column density in
systems for which \ion{O}{6} has been identified is 
$N({\rm HI})=6.6\times 10^{15}$~cm$^{-2}$, and so at $z=2.2$,
\begin{equation}
\Omega_{\rm g,OVI}\sim 6\times 10^{-3}h^{-1}
\end{equation}
For HE$2217-2818$ the total \ion{H}{1} column density which may be associated
with \ion{O}{6} systems is $1.9\times 10^{16}$~cm$^{-2}$. If this
quantity is more appropriate, then $\Omega_{\rm g,OVI}$ is three times
higher than the estimate above. Also, if the lower \ion{H}{1}
column density systems also contain \ion{O}{6} below the detection limit, then
the value will be higher. In any case this is a significant
fraction of the total baryon density $\Omega_b=0.02 h^{-2}$ inferred
from big bang nucleosynthesis \cite{OMeara01}. Hence, we conclude that
at $z\sim 2$ the \ion{O}{6} systems contain a cosmologically
significant fraction of both the metals and the baryons in the
universe.

\section{Conclusions}

Specific \ion{O}{6} absorption systems in the $2\simlt z\simlt 2.35$
Lyman absorption forests of QSOs HE$1122-1648$ and HE$2217-2818$ were
identified by self-consistently Voigt profile fitting the whole of the
available spectral range (with all higher order Lyman lines included).
If these two QSO sight lines are typical, then at redshifts $z\simlt
2.5$ \ion{O}{6} is usually present in absorption systems for which
$\log N({\rm HI}) \simgt 14$, This is similar to the column density
range for which \ion{C}{4} is detectable in spectra of $z>3$ QSOs
(e.g., Songaila \& Cowie 1996; Ellison et al.\ 2000).  Weak \ion{O}{6}
lines associated with lower column density Ly$\alpha$ systems may also
be present (Schaye et al.\ 2000), but are too difficult to detect
directly.  These observations indicate that \ion{O}{6} systems contain
a significant fraction of the baryons (and metals) in the universe.

For most systems the observed line widths, particularly those of the
Lyman lines, do not allow temperatures high enough that \ion{O}{6} may
be produced through collisional ionization. It is possible to contrive
a situation where the \ion{H}{1} in collisionally ionized \ion{O}{6} regions
is almost invariably blended with a cooler system with higher
\ion{H}{1} column density and no detected heavy elements, but even then
collisional ionization would require that all systems have nearly the
same unnatural temperature.

For most systems photoionization models with a substantial hard
ultraviolet flux reproduce the observations for expected densities and
heavy element abundances generally in the range $\sim
10^{-3}$-$10^{-2}$ solar.  These models give density estimates which
are in good agreement with those from the photoionized, local
hydrostatic equilibrium model of Schaye (2001). Assuming a hydrogen
ionization rate $\Gamma_{\rm HI} \approx 10^{-12}~{\rm s}^{-1}$ at
$2\simlt z\simlt 2.3$, the hydrogen densities are generally within
0.5 dex of $n_{\rm H}=10^{-4}$.

Photoionization by a UV spectrum much softer than that of Haardt \&
Madau (1996) ($\Gamma_{\rm HI} / \Gamma_{\rm HeII} \sim 10^2$) would
yield lower densities than would be expected for self-gravitating
clouds. However, this possibility is ruled out because it would result
in such large absorbers that the differential Hubble flow across the
absorbers would far exceed the observed line widths.

For some systems the interpretation is less straightforward. For
example, a high ionization system towards HE1122--1648 at redshift
$z=2.030$ appears to have heavy element abundances close to solar, so
is likely to be close to a metal-enriched galaxy.  Intriguingly,
regardless of whether the system is photoionized or collisionally
ionized, it can only be self-gravitating if its gas fraction is
negligible. Alternatively, it could be confined by external pressure or
it could be freely expanding.  In addition there are some absorption
systems for which the \ion{C}{4} and \ion{O}{6} redshifts clearly do
not agree well, and so there is some velocity separation of the
components in which these ions dominate. The system with the highest
\ion{H}{1} column density, at $z=2.180$ towards HE2217--2818, is the
best example. Similar systems with high \ion{H}{1} column density have
been  found by Simcoe et al. (2002). The relative frequency of
occurrence of the various systems remains a topic for further
investigation.

\section{\em Acknowledgments} 

We are grateful to the UVES P.I. Sandro D'odorico
and the UVES team for the excellent quality of the data delivered by
the instrument from the start of operation, and to Gary Ferland for
making freely available and supporting CLOUDY. Part of this work was
undertaken while RFC was supported by the ESO visitor program. JS is
supported by a grant from the W.~M.~Keck Foundation. This work has been
supported by the ``Physics of the Intergalactic Medium'' network set up
by the European Commission.

{}

\begin{table}
\caption{Voigt profile parameters for systems in HE1122--1648}
\label{tab:sample}
\begin{center}
\singlespace
\scriptsize
\begin{tabular}{lrrrrrr}
\tableline
\multicolumn{1}{c}{Ion}&\multicolumn{1}{c}{Redshift}&\multicolumn{1}{c}{$\pm$}&\multicolumn{1}{c}{$b$}&\multicolumn{1}{c}{$\pm$}&\multicolumn{1}{c}{$\log N$}&\multicolumn{1}{c}{$\pm$}\\ 
\tableline\tableline
\noalign{\smallskip}
\footnotesize
HI&2.100631&0.000014&20.84&1.64&13.392&0.097\\
HI&2.101378&0.000202&41.48& 11.19&14.462&0.351\\
HI&2.101764&0.000006&32.38&4.36&14.878&0.172\\
CIV&2.101764&&29.32&1.69&12.691&0.017\\
OVI&2.101764&&29.25&1.03&14.127&0.024\\
HI&2.102143&0.000068&24.64&3.48&14.030&0.323\\
HI&2.102733&0.000010&25.58&0.88&13.452&0.020\\
HI&2.104169&0.000234&32.19& 11.06&12.748&0.491\\
HI&2.104534&0.000026&29.18&1.38&13.554&0.077\\
HI&2.106061&0.000001&19.98&0.18&13.676&0.004\\
HI&2.106497&0.000048&151.55&6.70&13.368&0.017\\
HI&2.107762&0.000001&21.74&0.16&13.922&0.004\\
HI&2.109189&0.000005&15.96&0.75&12.685&0.017\\
HI&2.110505&0.000019&26.02&1.35&13.239&0.047\\
HI&2.110962&0.000017&29.62&1.50&13.409&0.032\\
HI&2.112169&0.000008&35.69&0.46&13.921&0.014\\
HI&2.112405&0.000004&15.57&0.76&13.505&0.033\\
HI&2.113518&0.000393&39.75& 23.21&12.879&0.580\\
HI&2.114065&0.000004&29.59&0.61&14.932&0.023\\
HI&2.115206&0.000002&32.58&0.21&14.879&0.010\\
HI&2.116542&0.000002&19.77&0.27&13.320&0.005\\
HI&2.117804&0.000006&22.81&0.89&12.797&0.014\\
HI&2.120450&0.000017&25.41&2.35&12.398&0.033\\
HI&2.121509&0.000008&36.72&1.13&12.972&0.011\\
HI&2.125407&0.000033&75.49&4.69&12.689&0.023\\
HI&2.128974&0.000002&20.01&0.22&13.433&0.004\\
HI&2.130186&0.000002&18.39&0.21&13.436&0.005\\
HI&2.130852&0.000006&32.07&1.00&13.143&0.010\\
HI&2.131843&0.000046&33.06&9.42&12.229&0.100\\
HI&2.132466&0.000004&21.43&0.58&13.245&0.017\\
HI&2.133359&0.000044&52.87&8.24&13.147&0.080\\
HI&2.133449&0.000008&24.10&2.76&12.992&0.137\\
HI&2.136924&0.000016&27.50&2.23&12.587&0.029\\
HI&2.142052&0.000005&25.05&0.70&12.928&0.10\\
HI&2.144270&0.000212&55.27&10.65&13.174&0.222\\
HI&2.144525&0.000012&23.53&0.89&14.065&0.94\\
HI&2.144809&0.000372&30.76&23.38&13.124&0.766\\
HI&2.145409&0.000012&20.41&1.13&12.995&0.46\\
HI&2.146187&0.000004&27.39&0.71&13.129&0.08\\
HI&2.146702&0.000014&13.99&1.95&12.168&0.56\\
HI&2.149243&0.000037&49.08&5.01&12.428&0.38\\
\tableline
\end{tabular}

\end{center}

Notes: Errors are $1\sigma$ estimates. The Doppler parameter $b$ is in
km~s$^{-1}$ and the ion column density $N$ is the number per
cm$^{-2}$.  Where no errors are given for the redshift its value is the
same as for the ion above with the same quoted redshift value. The full
table is available with the electronic version of this paper.
\end{table}

\begin{table}
\singlespace
\caption{\centerline{\ion{O}{6} systems in HE2217--2818 with $2.00<z<2.36$}}
\label{tab:OVI2217}
\begin{center}
\begin{tabular}{lcccccccc}
\tableline
Ion&Redshift&$\pm$&$b$&$\pm$&$\log N$&$\pm$&$b_{\rm turb}$&$T$(K)\\
\tableline\tableline
\noalign{\smallskip}
HI&2.074740&0.000004&28.07&2.08&14.890&0.036&$14.2\pm 0.9$&$3.6\pm 0.8\times 10^4$\\
CIV&&&15.87&1.11&12.887&0.015\\
OVI&&&15.48&0.69&14.046&0.155\\
HI&2.075398&0.000003&32.62&6.90&13.757&0.472&$5.9\pm 6.6$&$6.3\pm 3.0\times 10^4$\\
CIV&&&10.99&0.51&12.812&0.015\\
HI&2.075446&0.000006&27.94&3.83&13.891&0.290&$12.1\pm 1.20$&$3.9\pm 1.4\times 10^4$\\
OVI&2.075446&&13.62&0.87&14.301&0.029\\
\tableline
\noalign{\smallskip}
OVI&2.180424&0.000017&3.71&4.19&12.781&0.250\\
CIV&2.180488&0.000149&14.29&10.21&12.148&0.603\\
HI&2.180546&0.000382&27.33&9.06&14.576&1.296\\
OVI&2.180650&0.000027&25.81&2.93&13.723&0.049\\
CIV&2.180739&0.000026&13.93&1.61&13.348&0.093\\
SiIV&2.180746&0.000013&12.78&0.71&12.697&0.056\\
SiIII&2.180747&0.000011&12.42&0.65&12.889&0.054\\
HI&2.180791&0.000011&28.56&7.32&15.766&0.397\\
CII&2.180791&&10.36&1.68&12.732&0.050\\
SiIV&2.180838&0.000004&4.82&1.25&12.315&0.124\\
SiIII&2.180843&0.000006&4.77&1.34&12.485&0.108\\
CIV&2.180845&0.000009&7.00&2.75&12.682&0.385\\
CIV&2.181155&0.000031&12.78&4.44&12.184&0.122\\
SiIV&2.181174&0.000018&3.85&3.62&11.336&0.156\\
HI&2.181348&0.000356&49.59&11.52&14.841&0.320\\
OVI&2.181371&0.000022&14.34&3.29&13.228&0.079\\
CIV&2.181726&0.000016&8.87&1.52&12.450&0.102\\
OVI&2.181781&0.000015&9.30&2.14&13.416&0.134\\
SiIV&2.181804&0.000021&6.60&3.52&11.455&0.137\\
CIV&2.181917&0.000075&10.96&8.50&11.931&0.347\\
HI&2.181978&0.000060&23.15&4.51&13.973&0.263\\
OVI&2.182059&0.000030&19.67&3.61&13.723&0.072\\
\tableline
\end{tabular}

\end{center}

Notes: The turbulent component of the Doppler parameter $b_{\rm turb}$
and the temperature $T$ are evaluated using the $b^2=b^2_{\rm
turbulent}+b^2_{\rm thermal}$ model described in the text. Other details are as for Table \ref{tab:sample}.
\end{table}

\newpage
\begin{table}
\singlespace
\caption{\centerline{\ion{O}{6} systems in HE1122--1648 with $2.0<z<2.365$}}
\label{tab:OVI1122}
\begin{center}
\scriptsize
\renewcommand{\arraystretch}{0.8}
\begin{tabular}{lccccccll}
\tableline
\footnotesize
Ion&$z$&$\pm$&$b$&$\pm$&$\log N$&$\pm$&$b_{\rm turb}$&T\,(K)\\
\tableline\tableline
\noalign{\smallskip}
HI&2.007109&0.000004&29.46&1.40&15.433&0.099&$14\pm 1$&$4.1\pm 0.6\times 10^4$\\
CIV&&&15.87&0.47&13.100&0.014&&\\
SiIV&&&14.82&1.13&11.987&0.024&&\\
HI&2.007205&0.000005&23.50&2.49&15.135&0.302&$2\pm 5$&$3.4\pm 0.8\times 10^4$\\
CIV&&&6.98&1.61&12.309&0.072&&\\
OVI&&&6.10&1.60&13.424&0.080&&\\
\tableline
\noalign{\smallskip}
HI&2.030096&0.000003&22.96&51.2&11.738&0.300&$7\pm 10$&$0.0\pm 8.5\times 10^4$\\
CIV&&& 7.44&0.96&12.265&0.040&\\
NV &&& 7.44&0.57&12.848&0.021\\
OVI&&& 7.44&1.89&13.385&0.075\\
\tableline
\noalign{\smallskip}
HI&2.033133&0.000001&34.03&0.25&14.684&0.011&$15\pm 2$&$5.8\pm 0.3\times 10^4$\\ OVI&&&16.59&1.46&13.678&0.033&&\\
\tableline
\noalign{\smallskip}
HI&2.064412&0.000041&39.00&5.27&14.241&0.168&$39\pm 6$&$0.0\pm 4.0\times 10^4$\\
CIV&&&39.00&5.56&12.559&0.070&&\\
HI&2.064789&0.000016&72.50&21.46&14.052&0.420&$7\pm 17$&$3.2\pm 2.1\times 10^5$\\
CIV&&&21.91&2.19&12.520&0.070&&\\
HI&2.065681&0.000014&40.69&11.78&13.345&0.266&$13\pm 13$&$9.1\pm 6.3\times 10^4$\\
CIV&&&17.14&5.58&12.198&0.040&&\\
OVI&&&16.21&4.35&13.029&0.103&&\\
\tableline
\noalign{\smallskip}
HI&2.080162&0.000018&22.76&1.03&14.700&0.080&$20\pm 3$&$6.7\pm 7.0\times 10^3$\\
CIV&&&20.44&6.07&12.102&0.057 \\
OVI&&&20.38&2.39&13.503&0.054 \\
HI&2.080527&0.000021&17.74&1.40&14.236&0.154&$18\pm 3$&$0.0\pm 6.9\times 10^3$\\
CIV&&&17.74&8.39&11.864&0.093 \\
OVI&&&17.74&2.63&13.354&0.071 \\
\tableline
\noalign{\smallskip}
HI&2.101764&0.000006&32.38&4.36&14.878&0.172&$29\pm 1$&$1.3\pm 1.9\times 10^4$\\
CIV&&&29.32&1.69&12.691&0.017&&\\
OVI&&&29.25&1.03&14.127&0.024&&\\
 \tableline
\noalign{\smallskip}
HI&2.206460&0.000005&31.11&0.54&15.129&0.012&$7\pm 2$&$5.6\pm 0.3\times 10^4$\\
CIV&&&11.23&6.03&11.948&0.130&&\\
OVI&&&10.52&1.35&13.123&0.097&&\\
\tableline
\noalign{\smallskip}
HI&2.215271&0.000005&26.1&0.81&14.711&0.039&$3\pm 5$&$4.1\pm 0.3\times 10^4$\\
CIV&&&8.18&1.89&11.869&0.059&&\\
OVI&&&7.27&1.32&13.077&0.082&&\\
HI&2.215494&0.000031&38.40&3.60&14.335&0.157&$38\pm 10$&$0.3\pm 4.9\times 10^4$\\
CIV&&&37.78&6.95&12.113&0.064&&\\
OVI&&&37.77&9.34&13.351&0.164&&\\ 
\tableline
\noalign{\smallskip}
HI&2.339223&0.000154&23.81&2.90&13.959&0.943&&\\
HI&2.339260&0.000007&19.26&10.51&13.492&2.446&$8\pm 3$&$1.9\pm 2.6\times 10^4$\\
CIV&&&9.48&0.92&11.932&0.041&&\\
OVI&&&9.13&1.51&12.986&0.047&&\\
HI&2.339386&0.000283&24.34&9.21&13.793&1.060&&\\
\tableline
\noalign{\smallskip}
HI&2.352170&0.000005&21.14&0.78&14.415&0.038&$10\pm 1$&$2.1\pm 0.2\times 10^4$\\
CIV&2.352170&&11.15&0.79&12.383&0.024&&\\
OVI&2.352170&&10.81&0.82&12.770&0.126&&\\
CIV&2.352464&0.000007&7.53&1.13&12.150&0.058&&\\
HI&2.352695&0.000015&47.74&3.17&14.870&0.249&$47\pm 2$&$0.5\pm 2.3\times 10^4$\\
OVI&2.352695&&46.97&1.85&13.971&0.015&&\\
HI&2.352716&0.000061&57.30&7.96&14.549&0.505&&\\
CIV&2.352762&0.000005&16.66&0.86&12.800&0.020&&\\
CIV&2.353383&0.000016&33.96&2.53&12.667&0.027&&\\
\tableline
\end{tabular}
\renewcommand{\arraystretch}{1.0}

\end{center}
Note: Details as for Table \ref{tab:OVI2217}.
\end{table}

\begin{table}
\caption{\centerline{Photoionization models for HE$1122-1648$ systems}}
\label{tab:OVImod}
\begin{center}
\scriptsize
\begin{tabular}{ccccccccccc}
\tableline
\footnotesize
Redshift&\multicolumn{5}{c}{Log column density for:}&Model&&\\
$z$  &HI   &CIV  &NV &OVI  &SiIV &(see text)&$\log n_{\rm H}$&[C/H]&Temperature(K)\\ \tableline\tableline
2.080162&14.70&12.10&-&13.50&-&&-4.0&&$6.7\pm7.0\times 10^3$\\
\tableline
&14.70&12.10&12.25&13.50&\phantom{1}6.58&HM96&-4.15&-2.28&$5.1\times 10^4$\\
&14.70&12.10&12.27&13.50&\phantom{1}7.55&P1.5B0&-3.83&-2.14&$4.4\times 10^4$\\
&14.70&12.10&12.28&13.50&\phantom{1}7.20&P1.5B1&-4.79&-3.12&$6.0\times 10^4$\\
&14.70&12.10&12.26&13.50&\phantom{1}6.33&P1.5B2&-5.85&-4.13&$4.7\times 10^4$\\
&14.70&12.10&12.29&13.50&\phantom{1}7.85&P1.8B0&-4.10&-2.41&$4.7\times 10^4$\\
&14.70&12.10&12.30&13.50&\phantom{1}7.43&P1.8B1&-5.07&-3.40&$5.9\times 10^4$\\
&14.70&12.10&12.26&13.50&\phantom{1}7.43&P1.8B2&-6.19&-4.44&$4.1\times 10^4$\\
&14.70&12.07&12.50&13.53&10.32&Coll&  &-2.89&$2.5\times 10^5$\\
\tableline\tableline
2.080527&14.24&11.86&-&13.35&-&&-4.3&&$0.0\pm 6.9\times 10^3$\\
\tableline
&14.24&11.86&12.06&13.35&\phantom{1}6.11&HM96&-4.22&-2.00&$5.1\times 10^4$\\
&14.24&11.86&12.07&13.35&\phantom{1}7.07&P1.5B0&-3.91&-1.89&$4.4\times 10^4$\\
&14.24&11.86&12.08&13.35&\phantom{1}6.77&P1.5B1&-4.85&-2.86&$6.1\times 10^4$\\
&14.24&11.86&12.06&13.35&\phantom{1}5.91&P1.5B2&-5.91&-3.87&$4.6\times 10^4$\\
&14.24&11.86&12.08&13.35&\phantom{1}7.40&P1.8B0&-4.17&-2.17&$4.8\times 10^4$\\
&14.24&11.86&12.09&13.35&\phantom{1}7.02&P1.8B1&-5.14&-3.15&$6.0\times 10^4$\\
&14.24&11.86&12.06&13.35&\phantom{1}6.18&P1.8B2&-6.23&-4.16&$4.0\times 10^4$\\
&14.24&11.87&12.31&13.34&10.12&Coll&  &-2.63&$2.5\times 10^5$\\ 
\tableline\tableline
2.101764&14.88&12.69&-&14.13&-&&-3.9&&$1.2\pm 1.9\times 10^4$\\ 
\tableline
&14.88&12.69&12.86&14.13&\phantom{1}7.06&HM96&-4.19&-1.85&$4.9\times 10^4$\\
&14.88&12.69&12.88&14.13&\phantom{1}8.08&P1.5B0&-3.86&-1.72&$4.2\times 10^4$\\
&14.88&12.69&12.88&14.13&\phantom{1}7.68&P1.5B1&-4.82&-2.69&$5.9\times 10^4$\\
&14.88&12.69&12.86&14.13&\phantom{1}6.74&P1.5B2&-5.91&-3.72&$4.8\times 10^4$\\
&14.88&12.69&12.89&14.13&\phantom{1}8.36&P1.8B0&-4.14&-1.99&$4.6\times 10^4$\\
&14.88&12.69&12.89&14.13&\phantom{1}7.88&P1.8B1&-5.12&-2.99&$5.8\times 10^4$\\
&14.88&12.69&12.85&14.13&\phantom{1}6.87&P1.8B2&-6.23&-4.04&$4.2\times 10^4$\\
&14.88&12.69&12.90&14.13&\phantom{1}8.56&P2B0&-4.32&-2.17&$4.8\times 10^4$\\
&14.88&12.69&12.90&14.13&\phantom{1}8.03&P2B1&-5.31&-3.17&$5.8\times 10^4$\\
&14.88&12.68&13.11&14.14&10.92&Coll &  &-2.47&$2.5\times 10^5$\\ \tableline\tableline
2.206460&15.13&11.95&-&13.12&-&&-3.7&&$5.6\pm 0.3\times 10^4$\\
\tableline
&15.13&11.94&11.99&13.12&\phantom{1}6.89&HM96&-4.02&-2.98&$4.9\times 10^4$\\
&15.13&11.95&12.03&13.12&\phantom{1}7.94&P1.5B0&-3.49&-2.78&$4.1\times 10^4$\\
&15.13&11.95&12.04&13.12&\phantom{1}7.42&P1.5B1&-4.65&-3.77&$5.8\times 10^4$\\
&15.13&11.95&12.00&13.12&\phantom{1}6.43&P1.5B2&-5.73&-4.82&$5.0\times 10^4$\\
&15.13&11.95&12.06&13.12&\phantom{1}8.16&P1.8B0&-3.94&-3.03&$4.5\times 10^4$\\
&15.13&11.95&12.06&13.12&\phantom{1}7.61&P1.8B1&-4.94&-4.03&$5.8\times 10^4$\\
&15.13&11.96&12.40&13.11&11.18&Coll& &-3.50&$2.35\times 10^5$ \\
&15.13&11.95&11.61&13.12&\phantom{1}9.48&OP1.8B0&-3.39&-3.03&$3.4\times 10^4$\\
&15.13&11.96&11.68&13.12&\phantom{1}8.79&OP1.8B1&-4.44&-4.07&$5.0\times 10^4$\\
\tableline
\end{tabular}

\end{center}
Notes: The quantities in the rows in which a redshift is given are the
measured values, with the hydrogen density estimate $n_{\rm H}$
(cm$^{-3}$) and temperature derived as described in the text. The
values below those lines give the results of CLOUDY models, as
described in the text, in which the hydrogen density and heavy element
abundances relative to solar ([C/H]) were varied to provide a best fit
to the \ion{H}{1}, \ion{C}{4} and \ion{O}{6} column densities. The
temperatures there are the CLOUDY estimates. For collisional ionization
models (``Coll'' above) the temperature and abundances were varied to
get the best fit to the three column densities.
\end{table}

\begin{table}
\caption{\centerline{Voigt profile fits for collisional ionization models
with imposed high temperatures}}
\label{tab:OVIhiT}
\begin{center}
\begin{tabular}{lccccccll}
\tableline
Ion&Redshift&$\pm$&$b$&$\pm$&$\log N$&$\pm$&Temperature\,(K)\\
\tableline\tableline
\noalign{\smallskip}
HI&2.080268&0.000002&29.82&0.26&14.990&0.016&\\
HI &2.080152&0.000032&64.25&&&13.70&$2.5\times 10^5$\\
CIV&2.080152&&18.60&&11.684&0.309&\\
OVI&2.080152&&16.12&7.10&12.959&0.355&\\
HI &2.080384&0.000047&72.66&&&13.26&$2.5\times 10^5$\\
CIV&2.080384&&38.76&&12.240&0.101&\\
OVI&2.080384&&37.63&2.93&13.710&0.070&\\
 \tableline
HI&2.101668&0.000005&41.53&1.09&15.078&0.014&\\
HI&2.101769&0.000008&66.56&0.51&14.128&0.132&$2.5\times 10^5$\\
CIV&2.101769&&25.54&&12.666&0.020&\\
OVI&2.101769&&23.80&&13.875&0.069&\\
 \tableline
HI&2.206455&0.000007&29.26&0.53&15.144&0.018&\\
HI&2.206405&0.000012&62.27&&13.770&0.169&$2.35\times 10^5$\\
CIV&2.206405&&18.04&&12.451&0.062&\\
OVI&2.206405&&15.63&1.48&13.287&0.178&\\
\tableline
\end{tabular}

\end{center}
Notes: The temperature required to reproduce the \ion{C}{4}/\ion{O}{6} ratio was imposed on the \ion{H}{1} at that redshift. The associated \ion{H}{1} column density, and that of any new component required to fit the data are given. Other details as for Table \ref{tab:sample}.
\end{table}

\begin{table}
\caption{\centerline{Photoionization models for the $z=2.007$ components in HE$1122-1648$}}
\label{tab:OVImmx}
\begin{center}
\begin{tabular}{ccccccccccc}
\tableline
$z$  &HI   &CIV  &NV &OVI  &SiIV &Model&density&[C/H]&Temperature(K)\\ \tableline\tableline
2.007109&15.43&13.10&-&-&11.99&&-3.5&&$4.1\pm 0.6\times 10^4$\\
\tableline
&15.43&13.10&11.54&11.13&11.99&P1.8B0&-2.27&-1.16&$1.9 \times 10^4$\\
&15.43&13.10&10.92&\phantom{1}9.73&11.99&P1.8B1&-2.64&-1.61&$2.1 \times 10^4$\\
\tableline\tableline
2.007205&15.14&12.31&-&13.42&-&&-3.7&&$3.4\pm 0.8\times 10^4$\\
\tableline
&15.14&12.31&12.38&13.42&\phantom{1}8.63&P1.8B0&-3.80&-2.70&$4.3 \times 10^4$\\
&15.14&12.31&12.40&13.42&\phantom{1}8.07&P1.8B1&-4.80&-3.70&$5.7 \times 10^4$\\
\tableline
\end{tabular}

\end{center}
Note: Description as for Table \ref{tab:OVImod}
\end{table}

\begin{table}
\caption{\centerline{Models for the $z=2.030096$ system in HE$1122-1648$}}
\label{tab:OVImodX}
\begin{center}
\begin{tabular}{cccccccccc}
\tableline
$z$  &HI   &CIV  &NV &OVI  &SiIV &Model&$\log n_{\rm H}$&[C/H]&Temperature(K)\\ \tableline\tableline
2.030096&11.74&12.27&12.85&13.39&-&&&&$0.0\pm 8.5\times 10^4$\\
\tableline
&11.74&12.27&12.26&13.39&\phantom{1}7.82&HM96&-4.11&\phantom{-}0.69&$\phantom{5}2.0\times 10^4$\\
&11.74&12.27&12.30&13.39&\phantom{1}9.34&P1.8B0&-4.34&\phantom{-}0.70&$\phantom{5}5.0\times 10^3$\\
&11.74&12.27&12.33&13.39&\phantom{1}8.64&P1.8B1&-4.96&-0.30&$\phantom{5}2.1\times 10^4$\\
&11.74&12.26&12.71&13.41&10.49&Coll&&\phantom{-}0.20&$2.35\times 10^5$\\
\tableline
\end{tabular}

\end{center}
Note: Description as for Table \ref{tab:OVImod}
\end{table}

\end{document}